\documentclass[aps,prb,onecolumn, footinbib, 10pt]{revtex4-2}

\usepackage{dcolumn}
\usepackage{color}
\usepackage{makecell}
\usepackage{tabularx, graphicx}
\usepackage{latexsym}
\usepackage{colortbl}
\usepackage{psfrag}
\usepackage{bbm,bm,array,physics}
\usepackage{dsfont}
\usepackage{float, mathrsfs}

\usepackage{amsmath,amssymb} 
\usepackage{graphicx,comment}

\usepackage[tight]{subfigure} 

\usepackage[dvipsnames]{xcolor} 
\usepackage[papersize={8.5in,11in}]{geometry}
\usepackage[colorlinks=true]{hyperref}
\hypersetup{
    bookmarks=true,         
    unicode=false,          
    pdftoolbar=true,        
    pdfmenubar=true,        
    pdffitwindow=false,     
    pdfstartview={FitH},    
    pdfkeywords={keyword1} {key2} {key3}, 
    pdfnewwindow=true,      
    colorlinks=true,       
    linkcolor=magenta, 
    citecolor=blue,        
    filecolor=magenta,      
    urlcolor=blue           
} 

\def \nn{\nonumber \\}
\def\*#1{\boldsymbol{#1}} 

\begin{document}
\title{Fermi arcs for generic nodal points hosting monopoles or dipoles}

\author{Ipsita Mandal}
\email{ipsita.mandal@snu.edu.in}

\affiliation{Department of Physics, Shiv Nadar Institution of Eminence (SNIoE), Gautam Buddha Nagar, Uttar Pradesh 201314, India}

\begin{abstract}
Surface-localised states in the form of Fermi arcs emerge at the interfaces between three-dimensional topological semimetals and their surrounding vacuum, serving as primary signatures of momentum-space topology accessible to modern spectroscopic methods. These arcs encode information about topological singularities --- nodal points where multiple bands converge --- through their geometric properties. Specifically, the arcs terminate tangentially at the edges of Fermi-surface projections, delineating where surface modes transition into bulk extended states. The arcs' multiplicity at each node corresponds directly to the underlying topological charge (Berry-curvature monopole magnitude), thereby offering a window into momentum-space invariants via angle-resolved photoemission experiments. This work establishes a systematic framework for computing surface state configurations applicable to arbitrary nodal degeneracies and band dispersions: (1) nodes with twofold through multifold band multiplicities; and (2) bands exhibiting diverse momentum-dependence patterns, including linear, nonlinear, isotropic, and anisotropic regimes. We additionally examine topologically distinct nodal types where monopole charges vanish but dipolar Berry-curvature distributions appear, clarifying the surface-state landscape in such unconventional scenarios.
\end{abstract}

\maketitle

\tableofcontents

\section{Introduction}

\begin{figure}[t]
\centering
\subfigure[]{\includegraphics[width=0.35\textwidth]{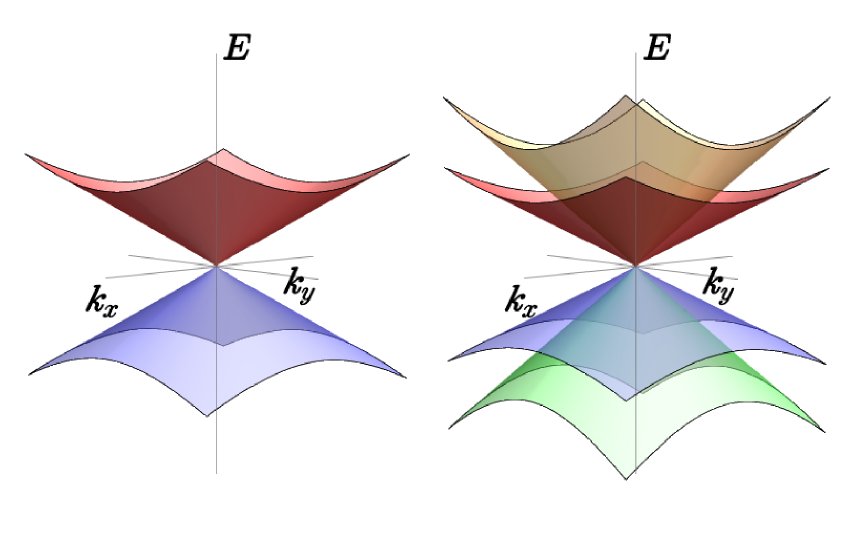}}\hspace{ 2 cm }
\subfigure[]{\includegraphics[width=0.45\textwidth]{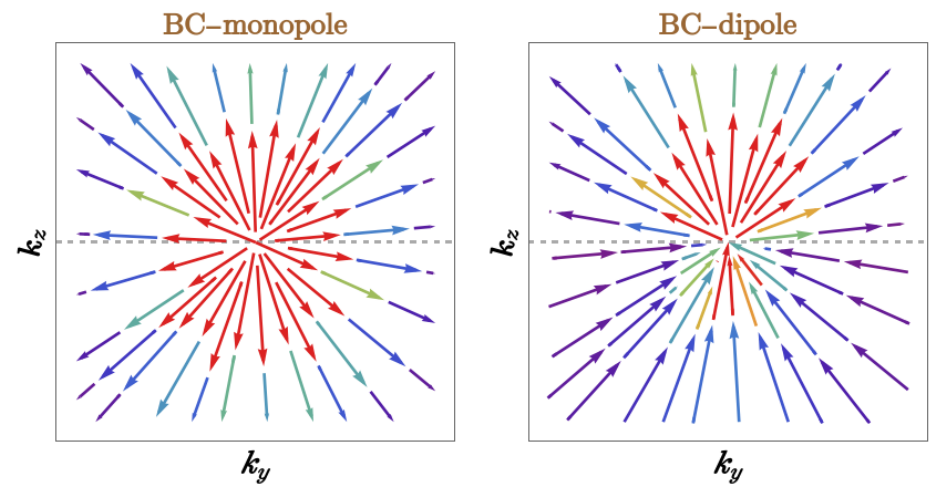}}
\caption{(a) Dispersion ($E$) in isotropic twofold (viz. Weyl node) and fourfold (viz. Rarita-Schwinger-Weyl node) band-crossings against the $k_x k_y$-plane. (b) Schematics of BC-flux lines in the $k_x = 0$ planes for a monopole and an ideal dipole. Here, the dipole-axis is orientied along the $k_z$-axis and, hence, the BC goes to zero on the $k_z=0$ plane in the 3d BZ.
\label{figdipole}}
\end{figure}

Three-dimensional (3d) semimetals containing degeneracies in the Brillouin zone (BZ) --- points where $(2\, \varsigma + 1)$ bands converge [cf.
Fig.~\ref{figdipole}(a)] --- offer a testbed for observing topological concepts embedded in condensed matter systems~\cite{burkov11_Weyl, hosur1, armitage_review, hosur-review, yan17_topological, ips-kush-review, bernevig, bernevig2, grushin-multifold, claudia-multifold}. Near a nodal point, the effective low-energy band structure follows the form
$ \boldsymbol d( \boldsymbol{k} ) \cdot \boldsymbol {S } $, with
$\boldsymbol d( \boldsymbol{k} ) = 
\lbrace d_x ( \boldsymbol{k} ) , d_y ( \boldsymbol{k} ) , d_z ( \boldsymbol{k} )  
 \rbrace.$
Here, $\boldsymbol k = \lbrace k_x, \,k_y, \,  k_z  \rbrace$ denotes momentum coordinates in the 3d BZ, and $\boldsymbol {S}$ is the vector operator with components $ \lbrace S_x, \,S_y, \,  S_z  \rbrace $ acting as angular-momentum operators. The band eigenvalues, spanning from $- \varsigma $ to $ \varsigma $, serve as pseudospin quantum numbers, distinguishing them from intrinsic electron spin.
An isotropic Weyl node in Weyl semimetals (WSMs)~\cite{burkov11_Weyl, armitage_review, yan17_topological} exemplifies the simplest case with $\varsigma = 1/2$. Higher pseudospin analogs have been discovered and characterized: triple-point semimetals (with $\varsigma = 1 $) and Rarita-Schwinger-Weyl (RSW) nodes ($\varsigma = 3/2$)~\cite{bernevig, long, isobe-fu,  tang2017_multiple, ma2021_observation, prb108035428, grushin-multifold, ips-rsw-ph, ips-shreya, ips-exact-rsw}. The topological character emerges from Berry-phase effects: closed-loop accumulation of geometric phases generates vector fields in momentum space, including Berry curvature (BC) and orbital magnetic moment (OMM). Nodes function as topological defects --- sources and sinks --- of the BC field [cf. Fig.~\ref{figdipole}(b)]. The BC singularity's strength is quantified by the Chern number ($\mathcal C$) of any closed oriented two-dimenionsal (2d) surface enclosing the node. Theoretical advances now extend beyond simple monopole configurations, revealing that nodes can exhibit higher-order multipolar character~\cite{twoband-dipole, graf-hopf, hopf2, hopf3} (such as dipoles and quadrupoles), echoing multipole hierarchies from classical electrostatics.

Momentum-space topology manifests in numerous experimentally measurable properties accessible through contemporary techniques. Transport measurements reveal a spectrum of topological responses with well-established theoretical underpinnings~\cite{timm, ips_rahul_ph_strain, rahul-jpcm, ips-ruiz, ips-tilted, ips-floquet, ips-kush-review, claudia-multifold, ips-rsw-ph, ips-shreya}. These encompass anomalous Hall conductivity~\cite{haldane04_berry, goswami13_axionic, burkov14_anomolous}, planar Hall phenomena \cite{zhang16_linear, chen16_thermoelectric, nandy_2017_chiral, nandy18_Berry, amit_magneto,  pal22a_berry, fu22_thermoelectric, timm, staalhammar20_magneto, yadav23_magneto, ips_rahul_ph_strain, rahul-jpcm, ips-ruiz, ips-tilted, ips-rsw-ph,ips-internode, ips-shreya, ips-spin1-ph, ips-exact-spin1, ips-nl-ph,  claudia-multifold, ips-exact-kwn}, Magnus forces~\cite{papaj_magnus, amit-magnus, ips-magnus}, optical dichroism \cite{ips-cd1, ips_cd}, and photocurrent generation \cite{moore18_optical, guo23_light,kozii, ips_cpge}. Yet among these signatures, Fermi arcs represent the most direct topological observable~\cite{yang_Weyl, lv_Weyl_arc, sanchez, spin1-arc,shroter-arc, Yao2020, multifold-rhsi, quad-weyl-arc}, being clearly visualized via angle-resolved photoemission spectroscopy (ARPES). Fermi arcs are surface-localised states in a 2d surface Brillouin zone (SBZ), geometrically constrained to terminate at the boundaries of bulk Fermi-surface (FS) projections, where they merge into extended bulk states.

Surface-localised Fermi arcs appear when considering semi-infinite slabs of topological semimetals with a surface normal $\boldsymbol{\hat n}$. These states emerge from the presence of topologically-protected edge modes in 2d momentum slices, which exist when open boundary conditions are imposed~\cite{burkov11_Weyl, hosur1, hosur-review}. While the momentum component parallel to the surface normal becomes ill-defined, the in-plane components $\boldsymbol k_\parallel$ perpendicular to $\boldsymbol{\hat n}$ retain their character as good quantum numbers in the large-slab limit. Consequently, a single interface defines an SBZ spanned by $\boldsymbol k_\parallel$ components, within which Fermi arcs reside. For Weyl semimetals, analytical surface-state solutions are well-established~\cite{witten, hashimoto-arc-weyl, babak-arc-weyl, inti-arc-weyl}. However, in more complex systems with multifold degeneracies or nonlinear dispersions, derivations remain largely case-by-case~\cite{witten, okugawa-arc-weyl, liu-arc-mutiweyl, jafari-arc-weyl, trauzettel-arc-weyl, ewelina-surface-states}, lacking systematic applicability across diverse dispersion landscapes and band configurations. This work develops a uniform computational scheme for deriving Fermi-arc configurations across arbitrary nodal geometries, establishing how these surface states encode the Chern numbers of underlying 2d momentum slices through bulk-boundary correspondence~\cite{burkov11_Weyl, hosur1, hosur-review}.

The paper is organised as follows: In Sec.~\ref{secbc}, we lay out our formulation of the boundary conditions, which leads to the solution for the surface states for a generic nodal-point semimetal. Sec.~\ref{secex} is devoted to the application of our method to diverse systems, spanning twofold to multifold ones, linear-in-momentum to quadratic- and cubic-in-momentum ones, and isotropic to anisotropic ones. Furthermore, we also consider two cases where the nodal points represent ideal BC-dipoles rather than monopoles. Finally, we end with a summary and outlook in Sec.~\ref{secsum}.

\section{Formulation of boundary conditions}
\label{secbc}

Let us take a single surface at $x=0$ such that the $x<0$ region represents a semi-infinite semimetal, with the bulk Hamiltonian $H(\boldsymbol k)$, where $\boldsymbol k \equiv \lbrace k_x, k_y, k_z \rbrace $ and $k =\sqrt{k_x^2 + k_y^2 + k_z^2}$. The translation symmetry is broken along the $x$-axis, and we use the Hamiltonian,
\begin{align}
H_x = H (k_x \rightarrow -i \, \partial_x , k_y, k_z)\,.
\end{align}
Demanding that $H_x $ be Hermitian in the region $x \in [0, \infty)$, we have the physical condition that
\begin{align}
\label{eqherm}
 \int_0^\infty  dx \,\psi^\dagger (x) \, H_x\,\phi (x) = 
 \int_0^\infty  dx\, [H_x\, \psi(x) ]^\dagger\,\phi(x) \,,
\end{align}
while considering two bonafide boundary-states, $\psi $ and $\phi$. The current-density operator is defined as
\begin{align}
j_x \equiv \partial_{k_x}  H(\boldsymbol k) \vert_{k_x \rightarrow -i\, \partial_x}\,.
\end{align}
Consequently, Eq.~\eqref{eqherm} also translates into a physically-sensible boundary condition (bc) which prohibits the current transmission through the boundary via $ (\psi^\dagger \, j_x \, \psi )\vert_{x=0}= 0$, alternatively known as the hard-wall bc. The boundary modes must behave as decaying wavefunctions of the form of
\begin{align}
\psi (x) = \psi_0 \, e^{- \kappa \, x} \text{ with } \text{Re}(\kappa) > 0\,,
\end{align}
such that the surface states are bound to the boundary at $ x= 0 $. Here, $\psi_0$ is independent of $x$ and any $x$-dependence of $\psi(x)$ is assumed to be contained in the $ e^{- \kappa \, x}$ factor. 

To avoid any confusion, we emphasise that $\psi(x)$ and $\phi(x)$ appearing in Eq.~\eqref{eqherm} denote two independent and arbitrary decaying functions on $[0,\infty)$. They are introduced solely as test functions in the conventional sense of self-adjoint-extension theory, as previously employed for WSM boundaries in Ref.~\cite{babak-arc-weyl}. Demanding the Hermiticity identity of Eq.~\eqref{eqherm} for arbitrary independent $\lbrace \psi,\phi \rbrace $ is what fixes the admissible, energy-independent boundary condition [see also Eq.~\eqref{eqalpha} below]. This defines a genuinely self-adjoint domain for $H_x$ on the half-space ($x<0$), analogous to fixing self-adjoint extensions of a differential operator via von Neumann's deficiency-index construction. Once this domain is fixed, any candidate surface-bound eigenstate must satisfy the same relation with $\psi$ set equal to itself, in order to belong to it. The resulting single-field current condition, $\psi_0^\dagger\, j_x\, \psi_0|_{x=0}=0$, used throughout this paper, is thus not a distinct or weaker requirement. It is precisely this general boundary condition evaluated on the specific physical solution at hand -- the standard hard-wall statement that the constructed surface state carries no probability current through the boundary.

\subsection{Witten's formulation for an untilted Weyl node}
\label{secwit}

The effective Hamiltonian in the vicinity of a single Weyl node [cf. Fig.~\ref{figdipole}(a)] is captured by
\begin{align}
\label{eqweyl}
H_W (\boldsymbol k) = \boldsymbol{\sigma} \cdot {\boldsymbol k}\,.
\end{align}
For this Hamiltonian, Eq.~\eqref{eqherm} translates into
\begin{align}
\label{eqjxweyl}
\psi^\dagger (x)\, \sigma_x \, \phi (x) \vert_{x=0} = 0
\Rightarrow j_x\vert_{x=0} \equiv \psi_0^\dagger\, \sigma_x \, \phi_0  = 0\,.
\end{align}

An energy-independent
boundary condition can be formulated as a local linear restriction on the components of the spinor wave function at the boundary, captured by \cite{witten}
\begin{align}
\label{eqalpha}
\Psi_0 =  M \, \Psi_0\,, \text{ where } M = \cos \alpha \, \sigma_y + \sin  \alpha \, \sigma_z \,.
\end{align}
This is a good boundary condition because $\sigma_x $ anticommutes with $M$, leading to
\begin{align}
 \psi_0^\dagger\, \sigma_x \, \phi_0 = 
 \frac{1}{2} \left[ \psi_0^\dagger\, \sigma_x \,M\, \phi_0 
 + ( M\, \psi_0)^\dagger\, \sigma_x \, \phi_0 \right ]
  = 0\,.
\end{align}
The variable $\alpha $ acts as a parameter and, by changing $\alpha$, we get Fermi arcs of various shapes and orientations with the restriction that they originate tangentially from the boundary of the projection of the FS on the $k_y k_z$-plane.

\subsection{Generic formulation applicable for any kind of nodes}
\label{secgen_method}

For generic nodes with multifold nodes and/or nonlinear powers of the components of $\boldsymbol k$, Witten's trick will not work, which we will explicitly see why considering some specific examples. Hence, we need an unambigous generic method applicable to derive the equations describing the curves representing the relevant Fermi arcs, which we describe here. Suppose we have an $N$-fold degeneracy at a node, described by the $N \times N $ Hamiltonian, $H_N (\boldsymbol k)$ in the bulk BZ. Let us look for surface states where the surface-normal is along the direction denoted by the unit vector, $\boldsymbol{\hat n}$, and located at $r_\perp = 0$, where $ r_\perp \equiv \boldsymbol r \cdot \boldsymbol{\hat n}$. We set our convention that the region $ r_\perp <0 $ represents the region occupied by the semimetallic material, while $r_\perp >0 $ represents the adjoining non-topological region (e.g., vacuum). Dividing up the momentum components along and perpendicular to $\boldsymbol{\hat n}$ as $  k_\perp$ and $\boldsymbol k_\parallel $, the boundary Hamiltonian is obtained as
\begin{align}
H_{r_\perp} = H_N ( k_\perp\rightarrow -i \, \partial_{r_\perp} , \boldsymbol k_\parallel)\,.
\end{align}
The imposition of the Hermiticity condition on $H_{r_\perp}$ leads to
\begin{align}
\label{eqherm1}
 \int_0^\infty  dr_\perp \,\psi^\dagger (r_\perp) \, H_{r_\perp}\,\phi (r_\perp) = 
 \int_0^\infty  dr_\perp\, [ H_{r_\perp}\, \psi(r_\perp) ]^\dagger\,\phi(r_\perp) \,,
\end{align}
analogous to Eq.~\eqref{eqherm}. Let us parametrise a solution as
\begin{align}
\label{eq11}
\psi (r_\perp) = \psi_0 \, e^{- \kappa \, r_\perp} \text{ with } \text{Re}(\kappa) > 0
\text{ and } 
 \psi_0^T  = \begin{bmatrix}
1 & a_1 +i \, b_1 & a_2 + i \, b_2 & \cdots & a_{N-1} + i \, b_{N-1}
\end{bmatrix}.
\end{align}
As argued earlier, (after some manipulations involving integration by parts to convert the integrands into total derivatives) Eq.~\eqref{eqherm1} will turn out to be equivalent to the condition of
\begin{align}
\label{eqcur}
\psi^\dagger  (r_\perp)\, j_{r_\perp} \psi (r_\perp) \big \vert_{r_\perp=0}= 0\,, \text{ where }
 j_{r_\perp} \equiv \partial_{k_\perp}  H_N(k_\perp,  {\boldsymbol k_\parallel} ) 
\big \vert_{k_\perp \rightarrow -i\, \partial_{r_\perp} }
\end{align}
represents the current-density operator perpendicular to the surface. Next, we need to solve for the eigenvalue equation,
\begin{align}
\label{eqhambdy}
H_{r_\perp} \psi (r_\perp) = E\, \psi (r_\perp)\,,
\end{align}
which leads to $N$ complex-valued equations involving the $(2N+1)$ unknown variables $ E$,
$\kappa_r \equiv \text{Re}(\kappa)$, $\kappa_i \equiv \text{Im}(\kappa)$, $a_1$, $ b_1$, $a_2$, $ b_2$, $\cdots$, $a_{N-1}$,
and $b_{N-1}$. This implies that we have $(2N+1)$ real equations [on including the real equation coming from Eq.~\eqref{eqcur}] at our disposal, which we can use to solve for the same number of unknown variables.
More details about intermediate steps can be found in the Appendix. There, we demonstrate explicitly that the general boundary term can be obtained by applying a Lagrange/Wronskian identity for higher-order derivative operators [cf. Eq.~\eqref{eqlagrange}] to each monomial order present in $H_N(k_\perp,\boldsymbol{k}_\parallel)$ in turn, and summing. The bulk contributions cancel identically, leaving the pure boundary functional $\Xi[\psi,\phi]$ [defined in Eq.~\eqref{eqxi}] evaluated at $r_\perp=0$, with the $r_\perp\to\infty$ contribution vanishing by the exponential decay of $\psi,\phi$. Specialising to the actual candidate surface state, $\phi=\psi=\psi_0\,e^{-\kappa \, r_\perp}$, and using that this same state solves the eigenvalue problem, $H_N(\lambda)\,\psi_0=E \, \psi_0$ with $\lambda\equiv i\,\kappa$ [cf. Eq.~\eqref{eqhambdy}], the boundary functional collapses -- exactly, and order by order in the momentum expansion of $H_N$, with no assumption that $\kappa$ is real -- onto $\Xi[\psi_0,\psi_0]|_{r_\perp=0}=\mathrm{Im}(E)\,|\psi_0|^2/\kappa_r$. Eq.~\eqref{eqherm1} is therefore equivalent, for the exponentially
decaying eigenstate ansatz of Eq.~\eqref{eq11} satisfying Eq.~\eqref{eqhambdy} and, for arbitrary finite polynomial order in
$k_\perp$, to the single real requirement that the surface-state energy be real [viz. $\mathrm{Im}(E)=0 $]. This is precisely the self-adjointness statement that a genuine surface-bound state must have a real eigenvalue. This reduces to the familiar single-field current condition of Eq.~(12), $\psi_0^\dagger\, j_{r_\perp}\,\psi_0|_{r_\perp=0}=0$, whenever $H_N$ is at most linear in $k_\perp$, as for the WSMs.
In the next section, we will study a variety of systems to see that the above procedure works generically.

\section{Specific examples}
\label{secex}

In this section, we will take some specific systems spanning generic features like nonlinear dispersion and multifold band-crossings. In the process, we will also observe how the emergent Fermi arcs reflect the underlying Chern numbers (or, equivalently, monopole charges) of the associated nodes. Additionally, we resolve the issue whether Fermi arcs can appear in nodes harbouring ideal BC-dipoles.

Before dealing with the individual systems, some comments are in order. For a non-interacting Hamiltonian in the vicinity of a nodal point, the relevant Matsubara Green's function is $G( i \, \omega,\boldsymbol{k})=[ i \, \omega - H_N(\boldsymbol{k})]^{-1}$, and the invariant protecting the point node against gap opening under continuous deformations is Volovik's $N_3$ \cite{volovik-book}, defined as the degree of the map from a three-sphere $S^3$ surrounding the singularity in $(\omega,\boldsymbol{k})$-space into the space of Green's functions,
\begin{align}
N_3 = \frac{1}{24\pi^2}\, \epsilon^{\mu\nu\rho\sigma}\, \mathrm{tr} \oint_{S^3} dS^\mu\; G\, \partial_\nu G^{-1}\, G\, \partial_\rho G^{-1}\, G\, \partial_\sigma G^{-1}\,.
\end{align}
The integration is performed over $S^3$.
For the effective low-energy Hamiltonians considered throughout the manuscript, $N_3$ reduces identically to the integral of the BC-flux over any Gaussian surface (e.g., $S^2$) enclosing the node, where the BC of the $s$-th band is
\begin{align}
\label{eqBCformula}
\boldsymbol{\Omega}_{s}(\boldsymbol{k}) = i\left[\nabla_{\boldsymbol{k}}\psi_{s}(\boldsymbol{k})\right]^\dagger \times \left[\nabla_{\boldsymbol{k}}\psi_{s}(\boldsymbol{k})\right].
\end{align}
Here, $\{ \psi_{s}(\boldsymbol{k}) \}$ is the set of normalised eigenvectors of the parent Hamiltonian. We can now use this to compute $N_3$ directly as the flux of $\boldsymbol{\Omega}_{s}(\boldsymbol{k})$ through any closed surface enclosing the node. Equivalently, this is the BC-monopole charge or Chern number that we invoke to count the number of Fermi arcs, and has already been computed over and over in the literature (e.g., Refs. \cite{ips-ruiz, ips-tilted, ips-cd1, ips-rsw-ph, ips-shreya, ips-internode, ips-spin1-ph, ips-exact-spin1, ips-exact-rsw, ips-nonlin-dipole}). The reduction from the frequency-momentum $S^3$-integral to the equal-time $S^2$-integral is standard for gapless free-fermion Green's functions and can be computed using Eq.~\eqref{eqBCformula}. Having made this identification explicit, we now tabulate $N_3$ for every variety of node discussed in this paper:
\begin{center}
\begin{tabular}{|c|c|c|}
\hline
Type of node & Band-multiplicity ($N$) & $ |N_3| = |\mathcal C| $ \\
\hline
 Untilted Weyl node & 2 & $ 1$ \\ \hline
Tilted Weyl node & 2 & $  1$ \\ \hline
Triple-point node & 3 & $  2$ \\ \hline
RSW node & 4 & $ 4$ \\ \hline
 Double-Weyl node & 2 & $ 2$ \\ \hline
Triple-Weyl node & 2 & $ 3$ \\ \hline
 Berry-dipole (two-band) & 2 & $0$ \\ \hline
Berry-dipole (three-band) & 3 & $0$ \\ \hline
 Quadruple-Weyl node & 2 & $  4$ \\
\hline
\end{tabular}
\end{center}
The two BC-dipole cases carry vanishing net monopole charge, $N_3=0$, since an ideal BC-dipole has equal and opposite monopole charge on its two poles. Nonetheless, as we discuss below, they still support Fermi arcs, since our boundary-condition construction is sensitive to more than the net charge alone.

\subsection{Untilted nodes of a Weyl semimetal}

\begin{figure}[t]
\centering
\includegraphics[width=0.75\textwidth]{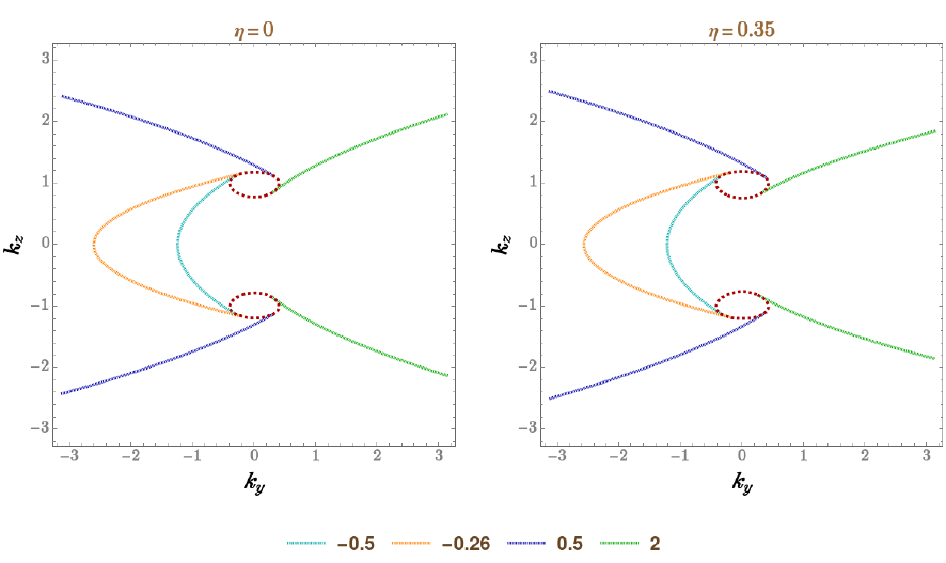}
\caption{A pair of conjugate Weyl nodes: The Fermi arcs represent modes with $E = \mu$, where $\mu =0.5$ and $k_w = 1 $. Each arc corresponds to a distinct value of $b_1$, colour-coded as shown in the plot-legends. The dashed red curves represent the projections of the bulk FSs on the $k_y k_z$-plane.
\label{fig_weyl}}
\end{figure}

Let us consider the one-node model given by Eq.~\eqref{eqweyl}, where we want to consider a boundary at $x=0$, indicating $\boldsymbol{\hat n} = \boldsymbol{\hat x} $. Using $ \psi_0^T  = \begin{bmatrix}
1 & a_1 +i \, b_1 \end{bmatrix}$, Eq.~\eqref{eqjxweyl} yields $ a_1  =  0 $. Next, the two components of the matrix equation in Eq.~\eqref{eqhambdy} take the forms of
\begin{align}
& b_1 \left(-i \,\kappa_i+k_y-\kappa_r\right)- E +k_z   = 0 
\text{ and }
-b_1 \left( E + k_z\right)+i \,\kappa_i+k_y+\kappa_r = 0 \nn
\Rightarrow & \, b_1 \left(k_y-\kappa_r\right)- E +k_z = 0\,, \quad 
b_1 \,\kappa_i = 0 \,, \quad \kappa_i = 0 \,, \text{ and }
-b_1 \left(E +k_z\right)+k_y+\kappa_r = 0\,,
\end{align}
after decomposing them into real and imaginary parts.
We obtain the solutions as $E = \frac{2 \,b_1\, k_y +(1- b_1^2)\, k_z} { 1 + b_1^2 }$, $\kappa_r 
= \frac{\left(b_1^2-1\right) k_y+2 \,b_1\, k_z} { 1 + b_1^2 } $, and $\kappa_i = 0$. We find that $b_1$ plays the role of a parameter, just like $\alpha$ in Eq.~\eqref{eqalpha}. The range of the Fermi arc is determined by the region where $\kappa_r > 0 $, i.e., $ \left(b_1^2-1\right) k_y+2 \,b_1\, k_z > 0$.  The arcs end tangentially on the projections of the bulk FSs, mixing with the bulk states, which is the point whose $k_y$- and $k_z$-coordinates are obtained as the solutions of the simultaneous equations, $E (k_y, k_z) =\mu$ and $\sqrt{k_y^2 + k_z^2} = \mu$. One can check that at this point, $\lbrace 
k_y, k_z \rbrace = \frac{\mu } { 1+b_1^2 } \lbrace 2\, b_1 , \, 1- b_1^2 \rbrace $ $\Rightarrow \kappa_r = 0 $. Therefore, the surface state ceases to exist at this point, merging with the bulk states.  

For the case of two nodes, separated along the $ k_z$-direction (for example) by an amount $2k_w$, we can easily get the nature of the Fermi arcs by the replacement $ k_z \rightarrow (k_z^2 - k_w^2)$. Fig.~\ref{fig_weyl} illustrates the Fermi arcs for four distinct values of $b_1$, when the chemical potential $\mu$ cuts the bandstructure of the surface bands.

\subsection{Tilted nodes of a Weyl semimetal}

The bulk Hamiltonian in the vicinity of a node, tilted with respect to the $k_x$-axis, is captured by
\begin{align}
\label{eqweylt}
H_W (\boldsymbol k) = \boldsymbol{\sigma} \cdot {\boldsymbol k} + \eta\, k_x \, {\mathbb{I}}_{2\times 2}\,.
\end{align}
For this case, on using $ \psi_0^T  = \begin{bmatrix}
1 & a_1 +i \, b_1 \end{bmatrix}$, Eq.~\eqref{eqcur} yields
\begin{align}
\psi_0^\dagger\, \sigma_x \, \psi_0 + \eta\,\psi_0^\dagger\,  \psi_0 = 0
\Rightarrow  2\, a_1 + \eta  \left( 1+ a_1^2+b_1^2 \right) = 0.
\end{align}
Next, the two components of the matrix equation shown in Eq.~\eqref{eqhambdy} take the forms of
\begin{align}
&  i \, \kappa_r \left(a_1+i \, b_1+\eta \right)-a_1 \kappa_i-i \,a_1 \,k_y-i \,b_1 \kappa_i
+b_1 \, k_y- E-\eta \, \kappa_i+k_z = 0 \nn
& \text{and }
i \left(i \, \kappa_i + k_y+\kappa_r\right)-\left(a_1+i \, b_1\right) 
\left[ E + \eta  \left(\kappa_i-i \,\kappa_r\right)+k_z\right ] = 0
\nn 
\Rightarrow & \,
-\left(a_1+\eta \right) \kappa_i+b_1 \left(k_y-\kappa_r\right)-E +k_z = 0\,, \quad 
-a_1 \,k_y+\left(a_1+\eta \right) \kappa_r-b_1 \, \kappa_i = 0\,, \nn
& \,-a_1 \left( E +\eta \, \kappa_i+k_z\right)-b_1 \, \eta \, \kappa_r-\kappa_i =0\,, \quad
a_1 \, \eta  \,\kappa_r-b_1 \left( E+\eta \, \kappa_i+k_z\right)+k_y+\kappa_r = 0\,.
\end{align}
We obtain the solutions as $E =  \frac{b_1 \left [\sqrt{1-\left( 1 + b_1^2 \right)\,
 \eta ^2}+1\right ] k_y+  \left[ \sqrt{1-\left( 1+ b_1^2 \right)\, \eta ^2}  - b_1^2 \right]
 k_z } { 1 + b_1^2} $, 
$\kappa_r  = \frac{\left(b_1^2-1\right) k_y+2 \,b_1\, k_z} { 1 + b_1^2} $, $\kappa_i = 0$, and $a_1 =
\frac{\sqrt{1-\left( 1 + b_1^2\right) \eta ^2} \, -1} {\eta }$.
While for the untilted case (i.e., $\eta = 0$), the projection is the locus of the curve $ k = \mu$ at $k_x = 0$, for the tilted case (i.e., $\eta \neq 0$), it is the locus of the curve $ k = \mu$ at $k_x = -\eta\, \sqrt{ (k_y^2 + k_z^2 ) / (1-\eta^2) }$ (or, equivalently, $k \,\sqrt{1-\eta^2} = \mu $ at $k_x = 0$). This is because, for $\mu \neq 0$, the anisotropic FS (shapeed like a spheroid elongated along the $k_x$-direction) has its largest radius of projection at a negative value of $k_x$. Consequently, the arcs end tangentially on the surface $\sqrt{
(k_y^2 + k_z^2)\,\sqrt{1-\eta^2} } = \mu $. Solving for  $k_y$ and $k_z$ using $E (k_y, k_z) =\mu$ in conjunction, we obtain the coordinates of the Fermi-arc's ending as $\lbrace k_y, k_z \rbrace 
= \frac{\mu} {\left( 1 + b_1^2 \right) \left( 1- \eta ^2\right)}
\left  \lbrace b_1 \left[ 1 + \sqrt{1-\left( 1+ b_1^2 \right) \eta ^2} \right ], 
 \, \sqrt{1-\left(b_1^2+1\right) \eta ^2} - b_1^2  \right \rbrace $, where of course $\kappa_r $ goes to zero. Again, for the case of two nodes, separated along the $ k_z$-direction (for example) by an amount $2k_w$, we can easily get the nature of the Fermi arcs by the replacement $ k_z \rightarrow (k_z^2 - k_w^2)$. Fig.~\ref{fig_weyl} illustrates the Fermi arcs for the two-node case using four distinct values of $b_1$, with $\eta$ set to $0.35$.

\subsection{Triple-point semimetal}

Let us now consider bands carrying pseudospin-1 quantum numbers crossing at threefold-degenerate nodes \cite{bernevig, spin13d1, ips-magnus, grushin-multifold, ips-spin1-ph,ips-exact-spin1, ips-internode}, representing the so-called triple-point semimetals (TSMs). These are three-band generalisations of the pseudospin-1/2 quasiparticles of WSMs, with the nodal points acting as BC-monopoles of magnitude 2. They can be realised in 3d tight-binding models for cold fermionic atoms in cubic optical lattices \cite{cold-atom} and, via \textit{ab initio} simulations, have been identified in materials like TaN, NbN, and WC-type ZrTe \cite{spin13d1, spin13d2, spin13d3, spin13d4}. The effective low-energy continuum Hamiltonian, in the vicinity of a threefold nodal point, is given by
\begin{align}
\label{eqham3d}
\mathcal{H}_{T}(\boldsymbol  k) = {\boldsymbol k} \cdot \boldsymbol{\mathcal S}  \,.
\end{align}
Here, $ \boldsymbol{\mathcal S} $ represents the vector spin-1 operator with three components,
\begin{align}
\mathcal{S}_x = \frac {1} {\sqrt{2}}
\begin{bmatrix}
0&1&0\\1&0&1\\0&1&0
\end{bmatrix} ,\quad
\mathcal{S}_y =\frac{1}{\sqrt{2}}
\begin{bmatrix}
0&- i &0\\ \mathrm{i} & 0 &-  i \\
0&  i  &0
\end{bmatrix} ,\quad
\mathcal{S}_z =
\begin{bmatrix}
1&0&0\\0&0&0\\0&0&-1
\end{bmatrix}.
\end{align}
The Chern-number values of $\pm 2$ indicate that we must have two distinct Fermi arcs associated with each node \cite{spin1-arc, shroter-arc}.

\begin{figure}[t]
\centering
\subfigure[]{\includegraphics[width=0.36\textwidth]{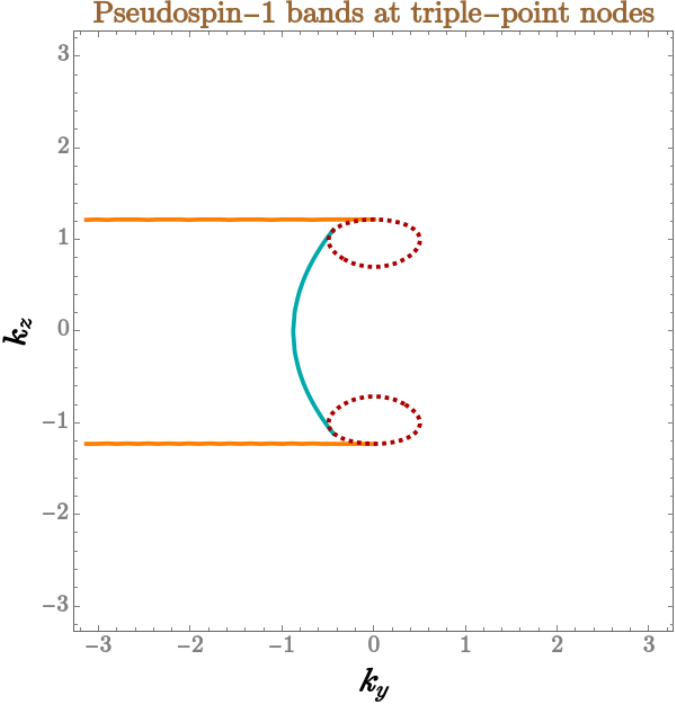}}\hspace{ 1 cm}
\subfigure[]{\includegraphics[width=0.365 \textwidth]{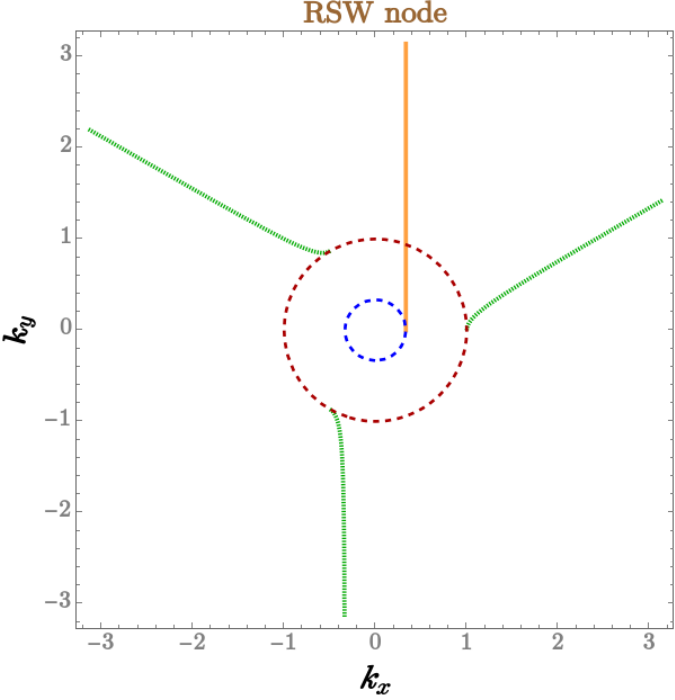}}
\caption{(a) A pair of triple-point nodes: The Fermi arcs represent modes with $E = \mu$, where $\mu = 0.5$ and $k_w =1$. The dashed red curves represent the projections of the bulk FSs (of the non-flat bands) on the $k_y k_z$-plane. The two Fermi arcs emanating tangentially from each FS-projection reflect the magnitude Chern number at each node equalling two. The cyan arc has been obtained by setting $b_1 =  -1$.
(b) RSW node: Four Fermi arcs emerging from the FS-projections of the two conduction bands, for $\mu=0.5$ and $b_1 = 0$, indicating the Chern number to be four.
\label{fig_spin1}}
\end{figure}

Here we note that, although this is a straightforward generalisation of the isotropic Weyl node, we cannot apply Witten's method  (outlined in Sec.~\ref{secwit}) simply because of the fact that $ \mathcal S_x$, $\mathcal S_y$, and $ \mathcal S_z $ do not anticommute (unlike the Pauli matrices). 
Furthermore, the boundary-condition matching procedure of Ref.~\cite{sukhachov-arc-spin1} relies on a heuristic model-specific reduction, namely, the boundary current happens to depend on only two of the three wave-function components for their particular choice of Hamiltonian. This renders their approach inadequate and non-generic for dealing with the present system.
Hence, our generic method works seamlessly to explicitly derive the appearances of the Fermi arcs. We use the parametrisation $ \psi_0^T  = \begin{bmatrix}
1 & a_1 +i \, b_1 & a_2 + i\, b_2 \end{bmatrix}$ and consider a boundary at $x=0$. Plugging it in Eq.~\eqref{eqcur} yields
\begin{align}
\psi_0^\dagger\, \mathcal S_x\, \psi_0 = 0 \Rightarrow a_1+ a_2 = 0\,.
\end{align}
Employing the matrix equation, viz. Eq.~\eqref{eqhambdy}, we get
\begin{align}
&  i \,\kappa_r -\sqrt{2} \left(a_1+i \,b_1\right) \left(E-k_z\right)  = 
\kappa_i+i\, k_y \,,\quad
-\left[ a_1+(1+i) a_2+i \,b_1\right) \left(\kappa_i-i \,\kappa_r\right]
+\left[i \,a_1+(1-i) \,a_2-b_1\right ] k_y
= \sqrt{2} \, E \,, \nn
& \text{and } \sqrt{2} \, (1+i) \,a_2 \left(E +k_z\right) =
 i \left(i \,\kappa_i+k_y+\kappa_r\right) \nn
& \Rightarrow \,\kappa_i = \sqrt{2} \,a_1 \left(k_z-E\right)\,,\quad
\sqrt{2} \,b_1 \left(k_z- E \right) +  \left(\kappa_r-k_y\right) = 0 \,, \quad
\left(a_1+a_2\right) \kappa_i+a_2 \left[ \kappa_r-k_y
+ b_1 \left(k_y+\kappa_r\right)\right ] = - \sqrt 2\,E \,, \nn &
\left(a_2+b_1\right) \kappa_i 
= a_2 \left(\kappa_r-k_y\right)+a_1 \left(k_y+\kappa_r\right), \quad
\kappa_i = -\,\sqrt{2}\, a_2  \left( E +k_z\right)\,, \quad
\kappa_r = \sqrt{2} \,a_2 \left( E +k_z\right)+k_y \,.
\end{align}
The solutions indeed tell us that there are two Fermi arcs characterised by (1) $E =\frac{2 \sqrt{2}\, b_1\, k_y+4 \,k_z}{ 2 + b_1^2 }-k_z$, $ \kappa_r = \frac{\left(b_1^2-2\right) k_y+2 \sqrt{2} \,b_1\, k_z} { 2 + b_1^2} $, $\kappa_i=0$, $a_1 = 0$, $a_2 = -\, b_1^2/2$, $|b_1 | > 0$, and $b_2 =0$; and (2)  $E = k_z$, $ \kappa_r = -\, k_y $, $\kappa_i=0$, $a_1 = 0$, $a_2 = 0$, $ b_1 = 0$, and $b_2 =0$. For the first arc, $b_1$ plays the role of a parameter and, hence, its shape changes as we vary $b_1$.
The points on the SBZ where $\kappa_r$ goes to zero are found to be $\lbrace k_y, k_z \rbrace =\frac{2\, \mu }{ 2 +b_1^2} \,\lbrace  
 b_1\, \sqrt{2}, \,b_1^2-2\rbrace $ (where $|b_1| > 0$) and $\lbrace k_y, k_z \rbrace = \lbrace  
 0, \,\mu \rbrace $, respectively, for the two Fermi arcs.

For the case of two nodes separated along the $ k_z$-direction (for example) by an amount $k_w$, which is possible when the nodes are not pinned at high-symmetry points for a given crystal structure, we can easily determine the nature of the Fermi arcs by the replacement $ k_z \rightarrow (k_z^2 - k_w^2)$. Fig.~\ref{fig_spin1}(a) illustrates the Fermi arcs for $\mu=0.5$, $k_w =1$, and $b_1=-1$. Since the nodes carry Chern numbers equalling $\pm 2$, we observe two arcs emanating from each FS-projection.

\subsection{Rarita-Schwinger-Weyl node in a chiral crystal}

Here we focus on bands carrying pseudospin-3/2 quantum numbers crossing at fourfold-degenerate degenracy-points [cf. Fig.~\ref{figdipole}(a)], widely known as the Rarita-Schwinger-Weyl (RSW) nodes. They appear at the $\Gamma$-points of chiral crystals with appreciable spin-orbit coupling (SOC) \cite{grushin-multifold, multifold-rhsi, Yao2020, ips-rsw-ph,ips-shreya,ips-internode}, hosting a net BC-monopole of magnitude 4. The effective Hamiltonian, in the vicinity of an isotropic RSW node, takes the form of
\begin{align}
\mathcal{H}(\boldsymbol{k}) =   {\boldsymbol k} \cdot \boldsymbol{\mathcal J} \,.
\end{align}
Here, $ \boldsymbol{\mathcal J } = \lbrace {\mathcal J }_x,\, {\mathcal J }_y,\, {\mathcal J }_z \rbrace $ represents the vector operator whose three components comprise the angular-momentum operators in the spin-$3/2$ representation of the SU(2) group. We choose the commonly-used representation with
\begin{align}
{\mathcal J }_x= 
\begin{pmatrix}
	0 & \frac{\sqrt{3}}{2} & 0 & 0 \\
	\frac{\sqrt{3}}{2} & 0 & 1 & 0 \\
	0 & 1 & 0 & \frac{\sqrt{3}}{2} \\
	0 & 0 & \frac{\sqrt{3}}{2} & 0 
\end{pmatrix} , \quad
{\mathcal J }_y=
\begin{pmatrix}
	0 & \frac{-i \,  \sqrt{3}}{2}  & 0 & 0 \\
	\frac{i \, \sqrt{3}}{2} & 0 & -i & 0 \\
	0 & i & 0 & \frac{-i \, \sqrt{3}}{2}  \\
	0 & 0 & \frac{i \, \sqrt{3}}{2} & 0 
\end{pmatrix}, \quad
{\mathcal J }_z =
\begin{pmatrix}
	\frac{3}{2} & 0 & 0 & 0 \\
	0 & \frac{1}{2} & 0 & 0 \\
	0 & 0 & -\frac{1}{2} & 0 \\
	0 & 0 & 0 & -\frac{3}{2} 
\end{pmatrix}.
\end{align}
Since the RSW node carries Chern number equalling $ 4 $, we must observe four arcs emanating from the FS-projections of the two bands being cut by $\mu$. For this case, just for the ease of calculations, we consider a boundary at $z=0$ (because the $\mathcal J_z$ matrix has fewer nonzero entries, leading to less cumbersome equations to be solved).
Here again we note that, although this is a fourfold generalisation of the isotropic Weyl node, we cannot apply Witten's method since $ \mathcal J_x$, $\mathcal J_y$, and $ \mathcal J_z $ \textit{do not anticommute} (unlike the Pauli matrices). Hence, we need to utilise our generic method to explicitly derive the equations of the Fermi arcs. We use the parametrisation $ \psi_0^T  = \begin{bmatrix}
1 & a_1 +i \, b_1 & a_2 + i\, b_2 & a_3 + i\, b_3  \end{bmatrix}$ because of the fourfold nature of the bands. Plugging it in Eq.~\eqref{eqcur} yields
\begin{align}
\psi_0^\dagger\, \mathcal J_z\, \psi_0 = 0 \Rightarrow 
3 -3\, (a_3^2 + b_3^2)+ a_1^2 +b_1^2 -(a_2^2+b_2^2) = 0\,.
\end{align}
Employing the matrix equation, viz. Eq.~\eqref{eqhambdy}, we get
\begin{align}
& 2\, E =\sqrt{3} \left(a_1+i \,b_1\right) \left(k_x-i \,k_y\right)
-3\, \kappa_i+3\, i \,\kappa_r\,,
\nn & \left(a_1+i \,b_1\right) \left(2 \, E +\kappa_i-i\, \kappa_r\right)
= \left(2 \,a_2+2 \, i \,b_2+\sqrt{3}\right) k_x
+\left( 2\, b_2-2 \,i \,a_2 +i \,\sqrt{3}\right) k_y\,,\nn
& \left(a_2 + i \,b_2\right) \left(2 \,E-\kappa_i + i \,\kappa_r\right)
=\sqrt{3}\, a_3 \left(k_x-i \,k_y\right)
+2\, a_1 \left(k_x+i \, k_y\right) + 2\, i \,b_1 k_x
+i\, \sqrt{3} \,b_3\, k_x - 2 \,b_1\, k_y+\sqrt{3}\, b_3 \,k_y\,, \nn
& \sqrt{3} \left(a_2+i \,b_2\right) \left(k_x + i \,k_y\right)
 = \left(a_3 + i\, b_3\right) \left(2 \, E- 3\, \kappa_i+3 \,i \,\kappa_r\right)
\nn & \Rightarrow 
2 \, E = \sqrt{3} \, a_1 \,k_x+\sqrt{3} \,b_1 \,k_y-3 \,\kappa_i \,,\quad
\sqrt{3}\, a_1 \,k_y = \sqrt{3} \,b_1 \,k_x + 3 \,\kappa _r \,,\quad
a_1 \left(2 \, E + \kappa _i\right) = \left(2 \,a_2+\sqrt{3}\right) k_x
+2\, b_2 \,k_y-b_1\, \kappa _r  \,,\nn
& \qquad 2 \, a_2\, k_y = a_1 \,\kappa _r-b_1 \left(2 \, E + \kappa _i\right)
+2\, b_2 \,k_x+\sqrt{3} \,k_y \,,\quad
2 \, b_1\,  k_y = a_2 \left(\kappa _i-2 \,  E \right)+2 \, a_1 \, k_x+\sqrt{3} \, a_3 \, k_x
+\sqrt{3} \, b_3 \, k_y+b_2 \, \kappa _r \,,\nn
& \qquad a_2  \,\kappa _r = 2  \,a_1 \, k_y-\sqrt{3}  \,a_3  \,k_y
+b_2 \left(\kappa _i-2  \, E \right)+2  \,b_1 \, k_x+\sqrt{3} \, b_3  \,k_x\,,\quad
2  \,a_3  \,E = 3  \,a_3 \, \kappa _i+\sqrt{3}  \,a_2  \,k_x-\sqrt{3}  \,b_2  \,k_y
+3  \,b_3  \,\kappa _r \,,\nn
& \qquad 2  \,b_3 \, E = \sqrt{3} \, a_2  \,k_y-3  \,a_3  \,\kappa _r
+3 \, b_3  \,\kappa _i+\sqrt{3}  \,b_2 \, k_x\,.
\end{align}
The explicit solutions are long and cumbersome. Hence, we do not write those explicitly. But, they can be parametrised in terms of one free parameter (which could be any of $\lbrace a_1, \, a_2, \, a_3, \,b_1, \, b_2, \,b_3 \rbrace $). In place of the explicit expressions, we show the arcs in Fig.~\ref{fig_spin1}(b) for one single case by setting $\mu=0.5$. The figure shows four Fermi arcs, in total, emanating outside the projection of the outer FS.

\subsection{Double-Weyl semimetal}
\label{sec-dws}

\begin{figure}[t]
\centering
\subfigure[]{\includegraphics[width=0.36\textwidth]{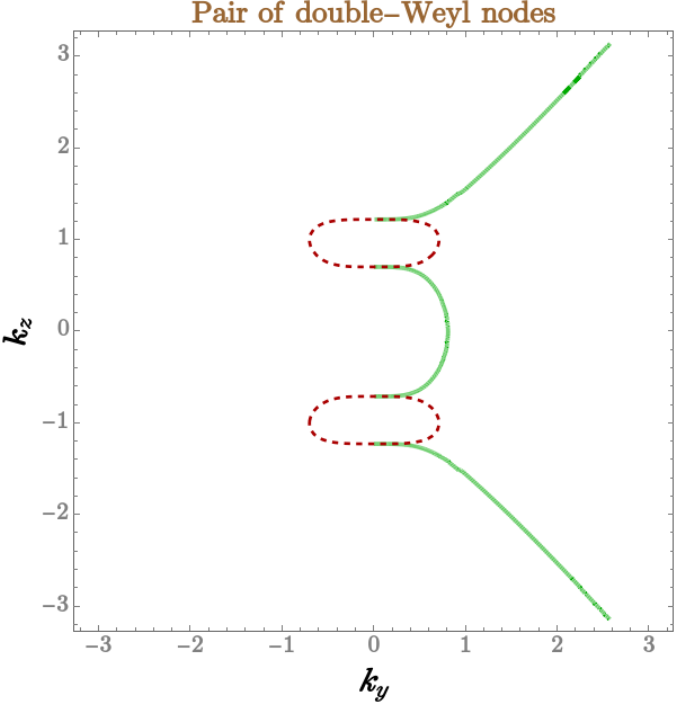}}\hspace{ 1 cm}
\subfigure[]{\includegraphics[width=0.36 \textwidth]{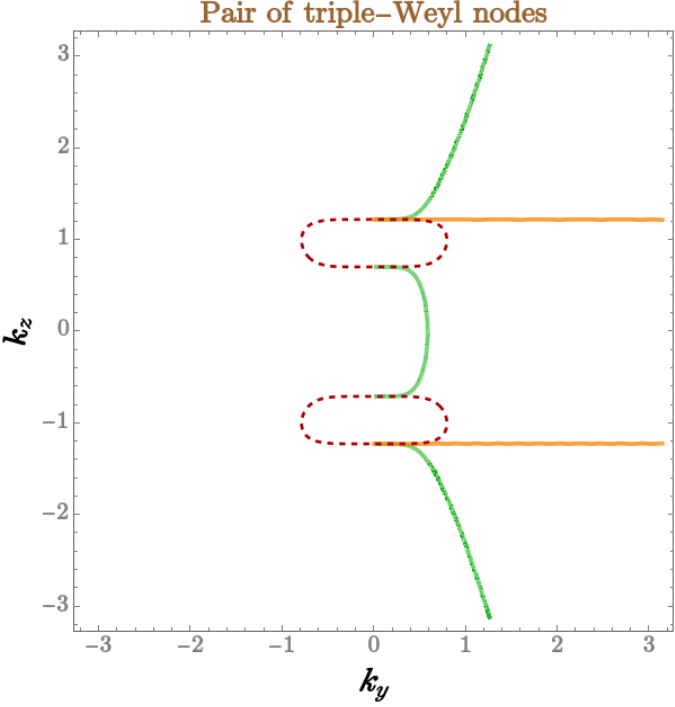}}
\caption{The Fermi arcs represent modes with $E = \mu$, where $\mu = 0.5$ and $k_w =1$. (a) A pair of double-Weyl nodes sources two arcs from the perimeter of each FS- projection. Here, $ t = \pi/6$ and $s=1$.
(b) A pair of triple-Weyl nodes sources three arcs from the perimeter of each FS-projection. Here, $ t = 0 $ and $s=1$.
\label{fig_multiweyl}}
\end{figure}

Double-Weyl nodes can be found in materials like $\mathrm{HgCr_2Se_4}$~\cite{Gang2011} and $\mathrm{SrSi_2}$~\cite{hasan_mweyl16, singh18_tunable}), which exhibit a hybrid of linear dispersion (chosen to be the $k_z$-axis) and quadratic dispersion (in the $k_x k_y$-plane). In the vicinity of such a nodal point, the low-energy effective continuum Hamiltonian is given by \cite{bernevig,bernevig2, fu22_thermoelectric, ips_rahul_ph_strain, ips-ruiz, ips-tilted, chen16_thermoelectric}
\begin{align}
\label{eqhamdw}
H_{dW} = \boldsymbol{d} \cdot \boldsymbol{\sigma}  \,, \text{ where }
\boldsymbol{d} \equiv \left\{ d_x,\, d_y, \, d_z \right\} = \left\{
\frac{k_x^2-k_y^2} {k_0} , \, \frac{2 \,k_x \,k_y} {k_0} ,\, k_z\right\} ,
\end{align}
where $k_0$ is a material-dependent parameter. For simplicity, we assume the momenta to be scaled such that $k_0$ can be set to 1.
Plugging in the parametrisation of $ \psi_0^T  = \begin{bmatrix}
1 & a_1 +i \, b_1 \end{bmatrix}$ in Eq.~\eqref{eqcur} leads to
\begin{align}
\label{eq29}
\kappa_i \,\psi_0^\dagger\, \sigma_x \, \psi_0 
=k_y \, \psi_0^\dagger\, \sigma_y \, \psi_0 \Rightarrow
 \kappa_i = \frac{b_1}{a_1} \,k_y \,.
\end{align}
Additionally, the two components of the matrix equation of Eq.~\eqref{eqhambdy} provide us with the following:
\begin{align}
&  E = k_z- \left(a_1+i\, b_1\right) \left(i \,\kappa _i-k_y+\kappa _r \right)^2
 \text{ and }
\left(a_1+i \,b_1\right) \left( E + k_z\right)+\left(i \,\kappa _i+k_y+\kappa _r\right)^2 = 0 \,.
\end{align}
The solutions are found easily after parametrising $ a_1 = a \cos t$ and $ b_1 = a \sin t$:
$E = \frac{\sqrt{k_z^2 \cos^2  t-4 \,k_y^4 \tan ^2 t }}
{\cos t}$, $ \kappa _r =  \frac{ s \, k_y}{\cos (t)} $, where
$a = s \, \frac{ k_z \,\cos t-\sqrt{k_z^2 \cos ^2 t-4 \,k_y^4 \tan^2 t} }
{2\, k_y^2 \left( 1 + \frac{1}{\cos t} \right)}$ and $s = \pm 1$.
Thus, $t$ serves an undetermined parameter, producing curves of varying shapes. The points on the SBZ where $\kappa_r$ goes to zero are found to be $\lbrace k_y, k_z \rbrace = \lbrace  0,\, \pm \mu \rbrace $.

For the case of two nodes separated along the $ k_z$-direction by an amount $k_w$, we can easily determine the nature of the Fermi arcs by the replacement $ k_z \rightarrow (k_z^2 - k_w^2)$. Fig.~\ref{fig_multiweyl}(a) illustrates the Fermi arcs for $\mu=0.5$, $k_w =1$, and $ t = \pi/6 $. Since the nodes carry Chern numbers equalling $\pm 2$, we observe two arcs emanating from the perimeter of each FS-projection.

\subsection{Triple-Weyl semimetal}

Analogous to the double-Weyl nodes, there exist triple-Weyl nodes (in materials like transition-metal monochalcogenides~\cite{liu2017predicted}) from which bands with anisotropic dispersion emerge. The triple-Weyl case comprises a hybrid of linear dispersion (chosen to be the $k_z$-axis) and quadratic dispersion (in the $k_x k_y$-plane). In the vicinity of such a nodal point, the low-energy effective continuum Hamiltonian is given by \cite{bernevig, bernevig2, fu22_thermoelectric, ips_rahul_ph_strain, rahul-jpcm, ips-ruiz, ips-tilted}
\begin{align}
\label{eqhamtw}
H_{tW} = \boldsymbol{d} \cdot \boldsymbol{\sigma}  \,, \text{ where }
\boldsymbol{d} \equiv \left\{ d_x,\, d_y, \, d_z \right\}=
 \left\{ \frac{k_x^3-3 \,k_x\, k_y^2} {k_0^2}, \, \frac{3 \,k_x^2 \,k_y-k_y^3} {k_0^2} ,
\, k_z\right\} \,,
\end{align}
where $k_0$ is a material-dependent parameter. Again, for simplicity, we set $k_0$ to 1.
Since $H_{tW}$ is cubic in $k_x$, Eq.~\eqref{eqcur} is understood here in the sense of the general boundary condition derived in the Appendix, $\Xi[\psi_0,\psi_0](0)=0$ [equivalently, $\mathrm{Im}(E)=0$], rather than as a literal evaluation of the current operator $j_{r_\perp}$ at the complex boundary-value $k_\perp\to\lambda$. As shown in the Appendix, the latter need not coincide with the correct condition beyond quadratic order in $k_\perp$. Plugging in the parametrisation of $ \psi_0^T  = \begin{bmatrix}
1 & a_1 +i \, b_1 \end{bmatrix}$ then leads to
\begin{align}
\left[ 3  \,k_y^2+\left(\kappa _r-i \,\kappa _i\right)^2
+\left(\kappa _r+i \,\kappa _i\right)^2-\kappa _r^2  - \kappa _i^2\right ]
\psi_0^\dagger\, \sigma_x \, \psi_0 
+ 6 \,\kappa _i k_y\, \psi_0^\dagger\, \sigma_y \, \psi_0= 0
\Rightarrow
a_1 \left( \kappa _r^2-3 \,\kappa _i^2+3 \,k_y^2 \right)+6\, b_1 \kappa _i \,k_y = 0\,.
\end{align}
Additionally, the two components of the matrix equation of Eq.~\eqref{eqhambdy} provide us with the following:
\begin{align}
&  E = k_z -\left(a_1+i\, b_1\right) \left(\kappa _i+i \,k_y-i \,\kappa _r\right)^3
 \text{ and }
i\, a_1 \left(e+k_z\right) =  b_1 \left( E +k_z\right)
+\left(i \,\kappa _i+k_y+\kappa _r\right)^3    \nn
\Rightarrow & \, E = k_z -a_1 \, \kappa _i 
\left [\kappa _i^2-3 \left(k_y-\kappa _r\right)^2\right ]
-b_1 \left(k_y-\kappa _r\right)
   \left [\left(k_y-\kappa _r\right)^2-3 \,\kappa _i^2\right ], \quad
b_1 \left( E +k_z\right)
=\left(k_y+\kappa _r\right) 
\left[ 3 \,\kappa _i^2 -\left(k_y+\kappa _r\right)^2\right ] ,\nn  
& \quad a_1 \left(k_y-\kappa _r\right) \left[\left(k_y-\kappa _r\right)^2-3 \,\kappa _i^2\right]
= b_1 \,\kappa _i \left[\kappa_i^2-3 \left(k_y-\kappa _r\right)^2\right ], \quad
\kappa _i \left [ 3 \left(k_y+\kappa _r\right)^2-\kappa _i^2\right ]
 =a_1 \left( E+k_z\right).
\end{align}
Finding exact analytical solutions is cumbersome because of the cubic roots involved. Instead, we parametrise $a_1 = a\cos t$ and $b_1 =\sin t$, and we show here the solutions for $t=0$:
(1) $E = k_z$, $\kappa_r  = \pm \, k_y$, $\kappa_i = 0$, and $a =  0 $;
(2) $ E = \frac{1}{3} \sqrt{9\, k_z^2-\frac{512 \,k_y^6}{3}}$, $\kappa_r = s \,k_y$, $\kappa^2 = 
\tilde s \,\frac{2 \, k_y }{ \sqrt 3}$, and $ a =  s\, \tilde s\,
\frac{3 \,\sqrt{3} \,k_z-\sqrt{27\, k_z^2-512\, k_y^6}}{64\, k_y^3}$, where $(s, \tilde s )= \pm 1$.
The points on the SBZ, where $\kappa_r$ for the three arcs goes to zero, are found to be $\lbrace k_y, k_z \rbrace = \lbrace  0,\, \pm \mu \rbrace $.

For the case of two nodes separated along the $ k_z$-direction by an amount $2k_w$, we can easily determine the nature of the Fermi arcs by the replacement $ k_z \rightarrow (k_z^2 - k_w^2)$.
Fig.~\ref{fig_multiweyl}(b) illustrates the Fermi arcs for $\mu=0.5$, $k_w =1$, and $s = 1$ (the value of $\tilde s $ does not matter) for the solution detailed above. Since the nodes carry Chern numbers equalling $\pm 3 $, we observe three arcs emanating from the perimeter of each FS-projection.

\subsection{Berry-dipole in a two-band model}

In a specific two-band model, the low-energy effective Hamiltonian in the vicinity of a single node harbouring a BC-dipole [cf. Fig.~\ref{figdipole}(b)] is captured by \cite{twoband-dipole}
\begin{align}
\label{eqbcd1}
H_{bd} (k_x, k_y, k_z)= 2\, v\, v_z \, k_z \left( k_x \, \sigma_x  +  k_y \,\sigma_y  \right )
+  \left [ v^2 \left( k_x^2 +k_y^2 \right) - k_z^2 \right ] \sigma_z \,.
\end{align}
Being an ideal dipole, this node carries vanishing BC-monopole charge. As such, one might think that there might be no Fermi arcs emerging on the SBZ. However, it turns out Fermi arcs do arise as surface states (on solving the relvant equations systematically using the generic method outlined in Sec.~\ref{secgen_method}), representing paired edge states on 2d slices.

\begin{figure}[t!]
\centering
\subfigure[]{\includegraphics[width=0.36\textwidth]{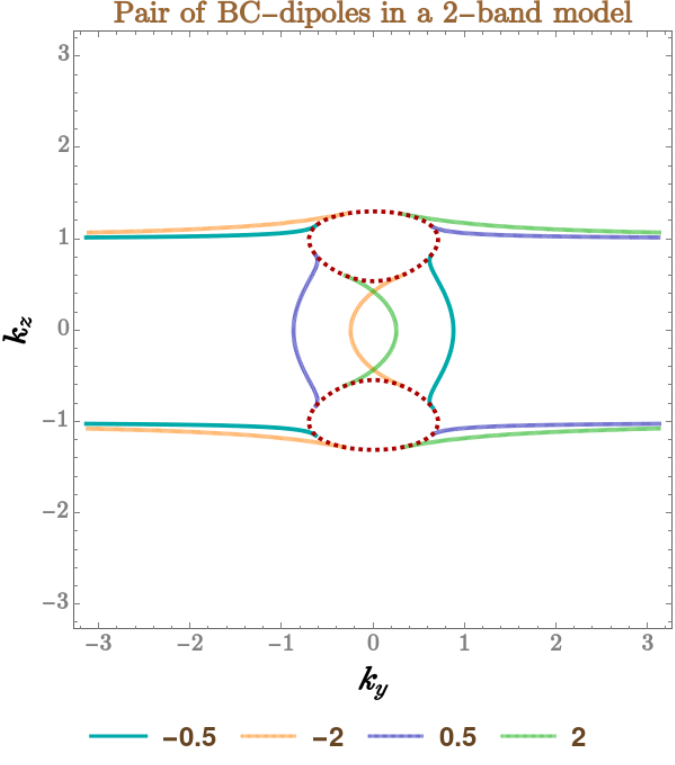}}\hspace{ 2 cm}
\subfigure[]{\includegraphics[width=0.36\textwidth]{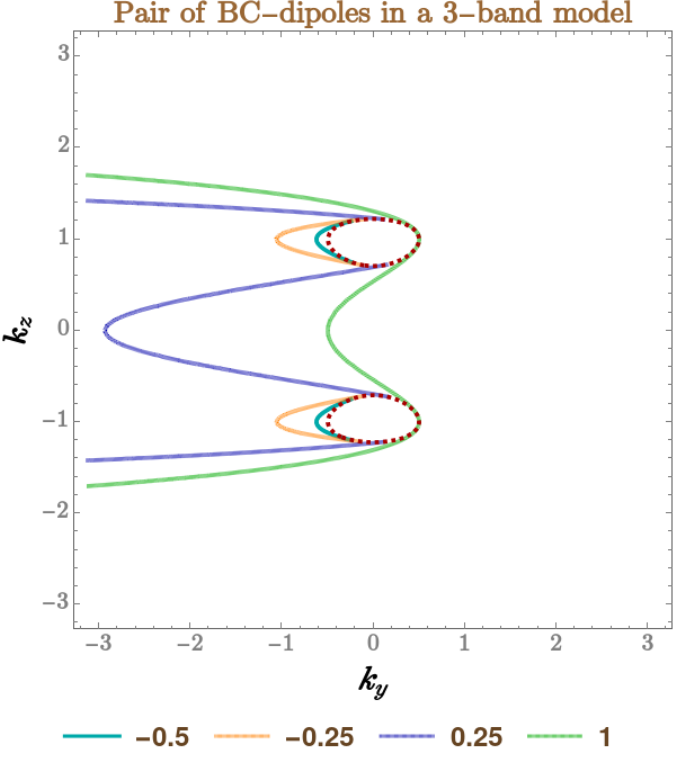}} 
\caption{A pair of nodes carrying ideal BC-dipoles using the model in (a) Eq.~\eqref{eqbcd1}, with $v_p= v_z=1$; (b) Eq.~\eqref{eqbcd2}.  In each case, Fermi arcs graze through each FS-projection at two distinct points, meeting tangentially along a FS-projection. The Fermi arcs represent modes with $E = \mu$, where $\mu =0.5$ and $k_w =1$. The dashed red closed contours represent the projections of the bulk FSs on the $k_y k_z$-plane. Each arc corresponds to a distinct value of $b_1$, colour-coded as shown in the plot-legends.
\label{fig_bcd1}}
\end{figure}

Since this is a two-band model, we again use $ \psi_0^T  = \begin{bmatrix}
1 & a_1 +i \, b_1 \end{bmatrix}$. Plugging it in Eq.~\eqref{eqcur} yields
\begin{align}
v_p^2 \,\kappa_i \,\psi_0^\dagger\, \sigma_z \, \psi_0 
+ v_p\, k_z\, \psi_0^\dagger\, \sigma_x \, \psi_0 = 0
\Rightarrow  2 \,a_1\, k_z- v_p  \left(a_1^2+b_1^2-1\right) \kappa_i = 0.
\end{align}
Next, the two components of the matrix equation of Eq.~\eqref{eqhambdy} take the forms of
\begin{align}
&  E =2 \left(b_1-i \,a_1\right)v_p \,v_z \, k_z 
\left ( k_y-\kappa_r  -\kappa_i \right)
+v_p^2 \left[ k_y^2-\left(\kappa_i+\kappa_r\right)^2 \right ]
-v_z^2 \,k_z^2  \nn
& \text{and }
 \left(a_1+i \,b_1\right) 
 \left [ v_z^2\, k_z^2  -v_p^2  \,k_y^2 + v_p^2  \left(\kappa_i+\kappa_r\right)^2  
 - E \right ]  +2\, i \,k_z\, v_p \,v_z 
\left(k_y+\kappa_r + \kappa_i \right)
= 0  \nn
\Rightarrow & \,
E = -2 \, b_1 \,v_p\, v_z \,k_z\left(\kappa_r+\kappa_i \right)
+2\, b_1 \,v_p\, v_z\,k_y\, k_z-v_p^2 
\left(\kappa_i+\kappa_r\right)^2+ v_p^2 \,k_y^2 -v_z^2  \,k_z^2\,,\quad
a_1 \, k_z  \left( \kappa_r + \kappa_i-k_y\right) =0\,,\nn
& a_1 \left(v_p^2 \left(\kappa_r + \kappa_i \right)^2-v_p^2 \, k_y^2 + v_z^2 \,k_z^2 
- E \right) = 0\,, \quad
b_1 \left[v_p^2 \left(\kappa_r + \kappa_i \right)^2
-v_p^2 \,k_y^2 +  v_z^2 \, k_z^2 - E\right ]
+2 \, v_p \,v_z\, k_z \left( \kappa_r + \kappa_i+k_y \right) = 0\,.
\end{align}
We obtain the solutions as $E =  \frac{k_z\, v_z}{b_1^2}
 \left[2 \,b_1 \,v_p \, k_y +\left(b_1^2-1\right) k_z \,v_z\right] $, 
$\kappa_r  =  k_y-\frac{k_z v_z}{b_1 v_p} $, $\kappa_i = 0$, and $a_1= 0 $.
The points on the SBZ where $\kappa_r$ goes to zero (and the arc becomes tangential to the projection of the FS) are found to be $\lbrace k_y, k_z \rbrace = \pm \,
\sqrt{\frac{\mu }{ 1 +b_1^2}} \,\lbrace  \frac{1}{v_p} , \, \frac{b_1}{v_z} \rbrace $, denoting antipodal points on a FS-projection. In fact, these tangents are inward and outward to the projection, being proportional to $\pm \lbrace  b_1, -1 \rbrace$. 
{\color{black}{
If the BZ contains two BC-dipole nodes, separated along the $ k_z$-direction by an amount $2k_w$, the answers are obtained by the replacement $ k_z \rightarrow (k_z^2 - k_w^2)$. Fig.~\ref{fig_bcd1}(a) shows the resulting Fermi arcs for a particular choice of the parameter-values. Evidently, two Fermi arcs graze past each FS-projection.
This is tied to a finer-grained momentum-resolved structure, obtained by treating each constant-$k_z$ slice of the BZ as an independent 2d Hamiltonian and examining its Chern number $\mathcal C(k_z)$. As shown in Ref.~\cite{twoband-dipole}, this layer Chern number vanishes identically for every $k_z$ lying between the two nodes. Because the Chern number is zero, no slice can host a single unpaired chiral edge mode --- as in a quantum-spin-Hall insulator, the boundary states are forced to appear as a counter-propagating pair. Unlike a genuine quantum-spin-Hall edge, however, this protection is furnished not by time-reversal symmetry, which is absent here, but by a chiral symmetry along a high-symmetry line of each $k_z$-slice, characterised by a $k_z$-dependent winding number. A consequence of this distinct protection mechanism is that the pair of Fermi arcs on a given surface is found to carry the same spin-polarisation, rather than the opposite spin-textures expected of a time-reversal-related pair, as would occur in a genuine quantum-spin-Hall edge or in the Fermi arcs of a Dirac semimetal.
}}

\subsection{Berry-dipole in a three-band model}

In Ref.~\cite{graf-hopf}, massless multifold Hopf semimetals (MMHSs) have been introduced which host BC-dipoles at nodes where linearly-dispersing bands cross. We consider the simplest case with a threefold-degenerate node, captured by the effective continuum Hamiltonian \cite{graf-hopf, hopf2, hopf3},
\begin{align}
\label{eqbcd2}
H_{H} = k_x \,\lambda_1 + k_y \,\lambda_2+ k_z\,\lambda_5\,,
\end{align}
where
\begin{align}
\lambda_1 =  \begin{bmatrix}
 0 & 1 & 0 \\
 1 & 0 & 0 \\
 0 & 0 & 0 \\
\end{bmatrix}, \quad
\lambda_2 = \begin{bmatrix}
 0 & -i & 0 \\
 i & 0 & 0 \\
 0 & 0 & 0 \\
\end{bmatrix}, \quad
\lambda_5 = \begin{bmatrix}
 0 & 0 & -i \\
 0 & 0 & 0 \\
 i & 0 & 0 \\
\end{bmatrix} .
\end{align}
For this 3-band model, we use the parametrisation $ \psi_0^T  = \begin{bmatrix}
1 & a_1 +i \, b_1 & a_2 + i\, b_2 \end{bmatrix}$. For a boundary at $x=0$, plugging it in Eq.~\eqref{eqcur} yields
\begin{align}
\psi_0^\dagger\, \lambda_1\, \psi_0 = 0 \Rightarrow a_1 = 0\,.
\end{align}
Employing the matrix equation, viz. Eq.~\eqref{eqhambdy}, we get
\begin{align}
& \left(b_2-i \,a_2\right) k_z+  b_1 \left( k_y -i  \,\kappa_i -\kappa _r\right) = E \,,
\quad k_y+\kappa_r + i \, \kappa_i = b_1 \, E \,, \text{ and }
E \left(a_2+i \,b_2\right)= i\, k_z \nn
\Rightarrow &\,  b_1 \left(k_y-\kappa_r\right)+b_2 \, k_z = E\,,\quad
a_2 \, k_z+b_1 \, \kappa_i = 0\,,\quad \kappa_i = 0 \,, \quad
b_1 \, E =  k_y+\kappa_r\,,\quad a_2\, E = 0\,,\quad b_2\, E = k_z\,.
\end{align}
The solutions turn out to be
$E = \frac{ b_1 \,k_y + \sqrt{b_1^2 \left(k_y^2+k_z^2\right)+k_z^2}} { 1 + b_1^2  }$,
$\kappa_r = \frac{ k_z^2 + b_1 \sqrt{b_1^2 \left(k_y^2+k_z^2\right) }-k_y}{ 1 + b_1^2}$, $\kappa_i  = 0 $,
$a_2 = 0 $, and $b_2  =  \frac{\sqrt{b_1^2 \left(k_y^2+k_z^2\right)+k_z^2}\,-b_1 k_y}{k_z}$. The points on the SBZ where $\kappa_r$ goes to zero (and the arc becomes tangential to the projection of the FS) are found to be $\lbrace k_y, k_z \rbrace = \mu \,\lbrace b_1 , \, \pm\, \sqrt{1-b_1^2}\rbrace $. This also tells us that the parameter $b_1$ is constrained to obey $|b_1|\leq 1 $. 
The corresponding tangents are proportional to $ \lbrace \pm\,\sqrt{1-b_1^2} , \, -b_1\rbrace $ and, hence, lead to the picture that one arc is entering and the other leaving the bulk-states. 
If the BZ contains two BC-dipole nodes, separated along the $ k_z$-direction by an amount $2k_w$, the answers are obtained by the replacement $ k_z \rightarrow (k_z^2 - k_w^2)$. 
Fig.~\ref{fig_bcd1}(b) demonstrates the nature of the Fermi arcs for some distinct values of $b_1$. Depending on the value of $b_1$, either two separate Fermi arcs graze past each FS-projection, or the same Fermi arc folds back and rejoins the same FS-projection. The overall behaviour is tied to the fact that the 2d layer-Hamiltonians for $k_z$ between $\pm k_w $ carry a vanishing Chern number. Yet they support a pair of counter-propagating (helical) edge modes -- one entering and one leaving -- much like the edge pair of a quantum-spin-Hall insulator. Outside this range, a 2d layer hosts no edge mode at all. The net chirality and, hence, the Chern number stays zero irrespective of whether zero, two, or four such edge-modes are present. A concrete grounding for the helical interpretation comes directly from the lattice model of Ref.~\cite{graf-hopf}. In their construction, $H_H$ arises as the critical point of a gapped tight-binding model, $h_N^{\text{Hopf}}(\mathbf{k})$. For this lattice model, the three weak topological invariants, namely the layer Chern numbers, are shown to vanish identically. Since a slice at fixed $k_z$ is one of these three orientations, this establishes directly that the layer Chern number of every $k_z$ slice is exactly zero, a property required for the Hopf-number classification to be well defined in the first place. At the same time, the BC vectors of the bands satisfy $\Omega_s = \Omega_{-s}$ [where $s \in (\pm 1, 0)$ is the band index]. Thus, the two dispersive bands carry BC of the same sign in contrast with a Weyl node (which features $\Omega_s = -\Omega_{-s}$ with $s=\pm 1$). Thus, our results coming from the continuum Hamiltonian $H_H$ are a defining consequence of the underlying chiral symmetry that these models possess. Because the layer Chern number is zero throughout the range of $k_z$ bounded by the two FS-projections, whatever edge states do appear must arise as a counter-propagating pair, in direct analogy with the edge pair of a quantum-spin-Hall insulator. Unlike the quantum-spin-Hall case, however, the protection here comes from the chiral symmetry of the Hopf Hamiltonian itself, rather than from time-reversal symmetry.

\subsection{Quadruple-Weyl node with monopole-charge 4}

A quadruple-Weyl node (QWN) with twofold degeneracy carries $|\mathcal C| = 4 $ \cite{charge4-1, charge4-2, charge4-3, charge4-4} and can exist only in spinless systems at specific time-reversal-symmetric points. Thus, they can appear in electronic bandstructures of materials with
negligible SOC or in the phonon spectra of artificial crystals (such as photonic crystals), all of which can be treated as spinless systems. Let us take the lattice model of Ref.~\cite{charge4-4}, where two nodes of opposite chiralities emerge at the $\Gamma$- and $R$-points. The node sitting at the $\Gamma$-point can be described by the Hamiltonian,
\begin{align}
\label{eqhamqw}
H_{qW} = \boldsymbol{d} \cdot \boldsymbol{\sigma}  \,, \text{ where }
\boldsymbol{d} \equiv \left\{ d_x,\, d_y, \, d_z \right\}= \left\{\frac{ v \left(k_x^2-k_y^2\right)}{2}, \,\frac{\left(k_x^2+k_y^2-2 \,k_z^2\right)}{2} ,
\,v\, k_x\, k_y\, k_z\right\} .
\end{align}
Here we fix $v$ to be positive. The energy eigenvalues are given by $\pm \sqrt{ v^2 \left [
2 \,k_x^2 \,k_y^2 \left(2 \,k_z^2-1\right)+k_x^4+k_y^4\right ]
+\left(k_x^2+k_y^2-2\, k_z^2\right)^2 } / 2$. Clearly, the dispersion of the bands is anisotropic, with a cubic dispersion along the $(111)$ direction and quadratic dispersion along any other direction. Here, we expect 4 Fermi arcs to emanate from the nodal point \cite{quad-weyl-arc}, which we will derive explicitly by considering a boundary at $x = 0$.

On using $ \psi_0^T  = \begin{bmatrix}
1 & a_1 +i \, b_1 \end{bmatrix}$, Eq.~\eqref{eqcur} yields
\begin{align}
\kappa_i \left( v\, \psi_0^\dagger\, \sigma_x \, \psi_0 + \psi_0^\dagger\, \sigma_y \, \psi_0 \right)
+ v \, k_y \, k_z \, \psi_0^\dagger\, \sigma_z \, \psi_0 = 0
\Rightarrow  2 \,\kappa_i \left(a_1 \,v+b_1\right) = 
v \left(1 - a_1^2-b_1^2\right) k_y \,k_z\,.
\end{align}
Employing the matrix equation of Eq.~\eqref{eqhambdy}, we get
\begin{align}
& a_1 \left [ (v-i) \left(\kappa_i-i\, \kappa_r\right)^2- (v+i)\, k_y^2
+2 \,i \,k_z^2\right ]
+b_1 \left[(1+i\,v) \left(\kappa_i-i\, \kappa_r\right)^2
+ (1-i \,v) \,k_y^2-2 \,k_z^2\right ]
=2 \,E +2 \, v \,k_y \,k_z \left(\kappa_i-i\,\kappa_r\right) \nn
& \text{and } a_1\, E = 
a_1 \,v \,k_y \,k_z \left(\kappa_i-i \,\kappa_r\right)
-i\, b_1 \, E + b_1\, v \,k_y\, k_z \left(\kappa_r+i \,\kappa_i\right)
+\frac{ v+i }{2}  \left(\kappa_i-i\, \kappa_r\right)^2-\frac{(v-i) \,k_y^2}{2} -i \,k_z^2
\nn & \Rightarrow
 2\, E + 2\, v \,\kappa_i\, k_y\, k_z
 = b_1 \left(\kappa_i^2+2\, v\,\kappa_i \,\kappa_r + k_y^2-2 \,k_z^2-\kappa_r^2\right)
 -a_1 \left[2 \,\kappa_i \,\kappa_r+v \left(\kappa_r^2-\kappa_i^2\right)+v \,k_y^2\right ],
\nn & 
\qquad a_1 \left(\kappa_i^2+2 \,v\, \kappa_i \,\kappa_r+k_y^2-2 \,k_z^2-\kappa_r^2\right) =
-\,b_1 \left(2 \,\kappa_i\, \kappa_r-v\, \kappa_i^2+v\, k_y^2
+v\, \kappa_r^2\right)+2 \,v \,k_y \,k_z\, \kappa_r\,,\nn & \qquad
2\,a_1 \, E =
2  \,a_1  \,v  \,\kappa_i  \,k_y  \,k_z+2  \,\kappa_r 
\left(b_1  \,v  \,k_y \, k_z + \kappa_i\right)+v  \,\kappa_i^2-v \, k_y^2-v \, \kappa_r^2 \,,\nn & \qquad
2 \,b_1 \, E =
-\,2 \,v\, \kappa_r \left(a_1\, k_y\, k_z+\kappa_i\right)
+2 \,b_1 \,v \,\kappa_i \,k_y\, k_z+\kappa_i^2+k_y^2-2\,k_z^2-\kappa_r^2 \,.
\end{align}
The admissible solutions for $\mu \geq 0 $ are: (1) $E = -\, \frac{v \left(k_y^2-k_z^2\right)}
{\sqrt{ 1 + v^2 }}$,
$\kappa_r = -\,\frac{v \,k_y\, k_z} {\sqrt{ 1 + v^2 }} $, $\kappa_i = \pm \, 
\sqrt{\frac{ 2 \,k_z^2 - k_y^2 \left[ v^2 \left(k_z^2-1\right) +1\right ]}
{ 1 + v^2}}$, $a_1= \frac{1}{\sqrt{v^2+1}}$, and $b_1  = \frac{-\, v}{\sqrt{v^2+1}}$; and
(2) $E =  \frac{v \left(k_y^2-k_z^2\right)}{\sqrt{ 1 + v^2 }}$,
$\kappa_r = \frac{v\, k_y\, k_z} {\sqrt{ 1 + v^2 }}$, $\kappa_i = \pm \, 
\sqrt{\frac{ 2 \,k_z^2 - k_y^2 \left[ v^2 \left(k_z^2-1\right) +1\right ]}
{ 1 + v^2}}$, $a_1= \frac{ -\,1}{\sqrt{v^2+1}}$, and $b_1  = \frac{v}{\sqrt{v^2+1}}$.

\begin{figure}[t!]
\centering
\subfigure[]{\includegraphics[width=0.3\textwidth]{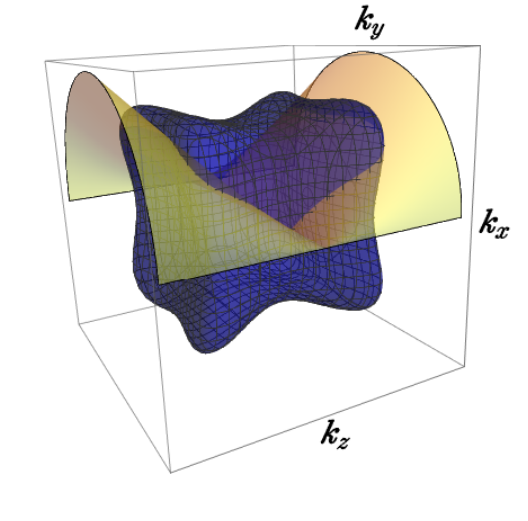}}\hspace{ 1 cm}
\subfigure[]{\includegraphics[width=0.6\textwidth]{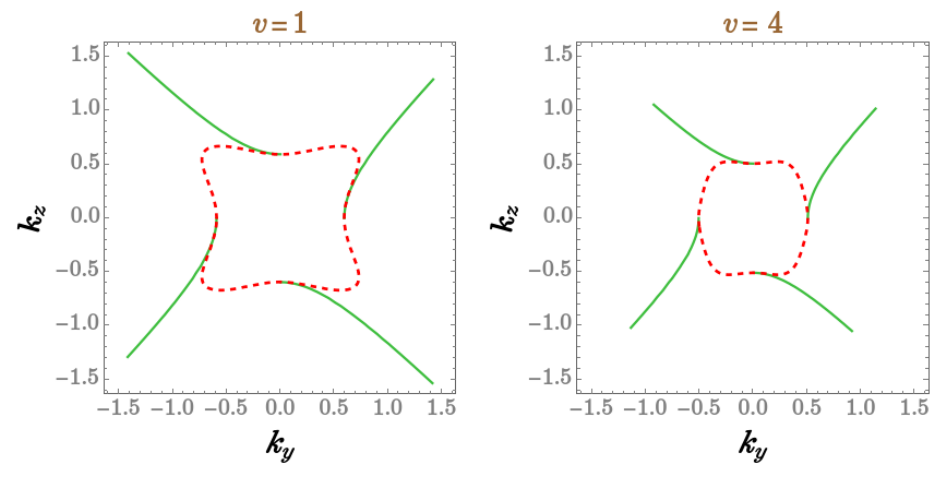}} 
\caption{(a) The FS-projection is obtained by cutting the 3d FS with the yellow-coloured 2d surface, captured by Eq.~\eqref{figanisofs}. Here, $\mu=0.25$ and $v = 1$. (b) The Fermi arcs represent modes with $E = \mu$, where $\mu =0.25$. The dashed red closed contour represents the projection of the bulk FS on the $k_y k_z$-plane.
\label{fig_qweyl}}
\end{figure}

Unlike the cases discussed so far, the projection of the FS on the $x=0$ boundary is not obtained by setting $k_x = 0$. Instead, it is given by the values $k_x = \pm \frac{\sqrt{ 2 \,k_z^2 -2 \,v^2 \,k_y^2\, k_z^2+ (v^2-1)\, k_y^2 }}
{\sqrt{ 1 + v^2 }}$, which leads to the closed curve (lying in the $k_y k_z$-plane) defined by 
\begin{align}
\label{figanisofs}
v\, \sqrt{\frac{ k_z^4- k_y^4 \left(k_z^2-1\right) \left( 1 + v^2 \,k_z^2 \right) + 2 \,k_y^2 \, k_z^2
   \left(k_z^2-1\right)  } { 1 + v^2 } }= \mu \,.
   \end{align}
The case of $v=1$ and $\mu=0.25$ is illustrated in Fig.~\ref{fig_qweyl}(a). Keeping this in mind, the points on the SBZ where $\kappa_r$ goes to zero, becoming tangential to the FS-projection, are found to be $\lbrace k_y, k_z \rbrace = \left\{0, \, \pm \frac{\sqrt{\mu } \,\sqrt[4]{ 1 + v^2}}{\sqrt{v}}\right\}$ and
$\lbrace k_y, k_z \rbrace =
\left\{ \pm \frac{\sqrt{\mu }\, \sqrt[4]{ 1 + v^2}}{\sqrt{v}} , \, 0\right\}$, respectively. Therefore, for different values of $v$, we get Fermi arcs of distinct orientations The tangents to the FS-projection, where the arcs meet the bulk states, are proportional to $ \left \lbrace \pm \,1, \, 0 \right \rbrace $ and $ \left \lbrace  0,\, \pm1 \right \rbrace $. These arcs are thus all leaving the bulk states at the projection.
Fig.~\ref{fig_qweyl}(b) illustrates the arcs for $\mu=0.25$ and two values of $v $. It might seem that the arcs end in thin air, but in actual lattice calculations, they will be found to end on some other nodal point or continue as arcs connecting with those ending on the conjugate QWN-projections located at the $R$-point. The effective Hamiltonian in Eq.~\eqref{eqhamqw}, correct only in the vicinity of the node, of course cannot capture such global details.

\section{Summary, discussions, and outlook}
\label{secsum}

From our analysis, we observe the following: While $N_3$ (and, hence, $\mathcal{C}$) is the topological invariant governing the stability of a point node itself, the existence and multiplicity of the associated Fermi arcs are more directly controlled by the Chern numbers $ \mathcal C(k_z)$ of the family of 2d Hamiltonians obtained by slicing the BZ at fixed values of one momentum component parallel to the surface (say, $k_z$). Nodes sharing the same value of $N_3$ are topologically equivalent in the sense of carrying the same number of protected arc branches, but generically differ in band-multiplicity and in the momentum-power/anisotropy of the dispersion, both of which are non-universal and set the detailed curvature of the arcs rather than their protected count. We provide below a comparison on a case-by-case basis:
\begin{itemize}

\item \textit{Untilted versus tilted Weyl node:} The tilt term $\eta\, k_x\, \mathbb{I}_{2\times2}$ existing in Eq.~\eqref{eqweylt} is proportional to the identity matrix and does not alter the eigenvectors of $H_W(\boldsymbol{k})$ at all. Thus, the associated BC and, hence, $N_3=\pm1$ is completely unchanged. Any visible difference of the resulting Fermi arcs in the two cases is thus not a topological effect at all, but it simply reflects the anisotropic reshaping of the bulk FS (and of its projection) by the tilt.

\item \textit{Triple-point node versus double-Weyl node:} Both carry $N_3=\pm2$ and, consistently, both source exactly two Fermi arcs per node. However, an TSM node is a threefold-degenerate isotropic crossing, while a double-Weyl node is a twofold-degenerate crossing with an anisotropic dispersion. The value of $N_3$, being a total flux (i.e., an integral quantity), is insensitive to this difference in band-multiplicity and angular/radial distribution of the underlying BC --- it only counts the net number of protected chiral branches guaranteed by bulk-boundary correspondence, and not their microscopic origin. This is precisely why the two arc families look qualitatively different. The detailed curvature of an individual arc is controlled by non-topological data [e.g., the local dispersion and the boundary-condition parameter(s)], rather than by $N_3$ itself.

\item \textit{RSW node versus quadruple-Weyl node:} Both carry $N_3=\pm4$ and both source four Fermi arcs, yet the RSW node is a fourfold-degenerate fully-isotropic linear crossing, whereas the QWN is twofold-degenerate with nonlinear anisotropic dispersion. Correspondingly, the bulk FSs and, thus, the shapes of their projections onto the SBZ, are entirely different in the two cases. In the QWN case, the projected FS is not even a simple curve of the form $k=\mu$, but instead is a more intricate closed contour. Once again, $N_3$ dictates only the topologically-protected count of arcs (four, in both cases), while their detailed geometry is fixed by the non-universal bandstructure.

\item \textit{2-band or 3-band BC-dipole node:} Here $N_3=0$, yet a pair of Fermi arcs is found to graze each FS-projection. The existence of the arcs is tied to a finer-grained momentum-resolved structure, obtained by treating each constant-$k_z$ slice of the BZ as an independent 2d Hamiltonian and examining its Chern number $\mathcal C(k_z)$. As shown in Ref.~\cite{twoband-dipole}, this layer Chern number vanishes identically for every $k_z$ lying between the two nodes. Because the Chern number is zero, no slice can host a single unpaired chiral edge mode. Therefore, as in a quantum-spin-Hall insulator, the boundary states are forced to appear as a counter-propagating pair. Unlike a genuine quantum-spin-Hall edge, however, this protection is furnished not by time-reversal symmetry, but by a chiral symmetry.

\end{itemize}
To summarise, nodes sharing the same $N_3$ are, from the point of view of the bulk topological classification alone, indeed indistinguishable. After all $N_3$ is a coarse integer invariant that only counts the total winding of the Green's function (equivalently, the net Berry flux) around the node, and says nothing about how that total charge is distributed among the internal (pseudospin/band) degrees of freedom, or (for the dipole cases) about how it is distributed in $k_z$. Our boundary-condition formalism, however, is sensitive to more than $N_3$ alone: it depends on the full polynomial structure of $H_N$. In particular, it depends on the band multiplicity $N$ and on the power(s) of momentum through which the monopole charge is realised. Consequently, they yield Fermi arcs of visibly-different shapes, curvature, and parameter dependence, even though the number of arcs terminating at each FS-projection (fixed by $|N_3|$, where non-zero) is the same.

It is instructive to compare the arc families derived above with Fermi arcs resolved by ARPES. The number of arc branches emanating from a given nodal projection is fixed by the topological invariant $N_3$ of the node (equivalently, by the Chern number). ARPES measurements on WSMs of the TaAs family reveal short Fermi arcs consistent with $N_3=\pm1$. Chiral crystals such as CoSi, RhSi, and PtGa, on the other hand, host multifold nodal degeneracies (or higher-charge nodes), exemplified by our RSW and pseudospin-1 models. These display long Fermi arcs that traverse an appreciable fraction of the SBZ and connect nodes of opposite chiralities located at different high-symmetry points \cite{takane2019_observation,multifold-rhsi,mengyu}, consistent with $|N_3|=4$ obtained here. Both features are captured naturally by our results: the number of arc branches sourced at a given projected FS is fixed by $N_3$, exactly as observed --- single arcs for TaAs-like materials and multiple/longer arcs for the higher-charge multifold semimetals. This protected count is entirely fixed by bulk topology and is insensitive to the boundary-condition parameters. By contrast, the pronounced curvature, kinking, and sometimes counter-intuitive (``bizarre'') shape of individual arcs seen within a single ARPES data set, even for the same bulk charge, is controlled by our boundary-condition parameters in the notation of Sec.~\ref{secex}. These parameters are fixed by the microscopic details of the surface termination and, therefore, expected to vary between samples, cleaves, or materials sharing the same bulk topology. In other words, the qualitative topologically-protected features of Fermi arcs (their number and connectivity to bulk FSs' projections) are governed by $N_3$. Their sample-dependent appearance in ARPES reflects the family of admissible boundary conditions instead, which are a direct symmetry-allowed consequence of surface reconstruction/termination chemistry. Our generic construction shows explicitly how a topologically fixed number of arcs can nonetheless be continuously deformed in shape by these parameters. This offers a concrete mechanism for the diversity of arc geometries reported experimentally for materials sharing the same bulk nodal charge, as well as parametrising this correspondence explicitly and analytically for the first time across a range of nodal-point types.

Our derivations for the nature of surface states in 3d semimetals provide an exhaustive analysis of generic boundary conditions for open boundaries. The main line of argument hinges on the preservation of self-adjointness (i.e., Hermiticity) of the effective Hamiltonian of the corresponding surface, which must host eigenstates that decay exponentially as we move away from the boundary with the topological material. On the side of the material itself, the states must merge with the eigenstates (of the bulk Hamiltonian) representing the 3d bulk states. Through explicit examples, we have shown how the number of Fermi arcs, arising in the SBZ, gives us a clear signature of the net Chern number at a given nodal point, conforming to the well-established notion of bulk-boundary correspondence. However exotic the bandstructure might be, our formalism will work in all generic cases. In the future, we plan to investigate the dynamics of the surface states, for example when a magnetic field is applied, when we can observe quantum oscillations \cite{qtm-osc-ashvin, hosur-arc-osc}.


\section*{Acknowledgments}

This research, leading to the results reported, has received funding from the Council of Science \& Technology (CST), U.P.

\appendix

\section*{Appendix: Derivation of the boundary conditions}

{\color{black}{In this appendix, we demonstrate explicitly that the general boundary term can be obtained by applying a Lagrange/Wronskian identity for systems containing generic higher-order-derivative operators. We break the underlying derivation up into the following steps:
\begin{itemize}

\item Step 1 (set-up): 
\\We write the bulk Hamiltonian as a matrix polynomial in the perpendicular momentum as
$$H_N(k_\perp,\boldsymbol{k}_\parallel)=\sum_{n\geq 0} C_n(\boldsymbol{k}_\parallel)\, k_\perp^n\,.$$
Hermiticity of $H_N$ for every real $k_\perp$ requires each $C_n$ to be a Hermitian matrix. Throughout the derivation, $k_\perp$ should be understood as the differential operator, $-i\, \partial_{r_\perp} $.

\item Step 2 (a Lagrange/Wronskian identity for a single power of $k_\perp$):
\\By induction on the degree $n$, one establishes the operator identity
\begin{align}
\label{eqlagrange}
\psi^\dagger(r_\perp) \, \big[k_\perp^n \, \phi(r_\perp)\big ]
 - \big [k_\perp^n \,\psi(r_\perp)\big]^\dagger \,\phi(r_\perp)
= - \, i\, \partial_{r_\perp} 
\sum_{m=0}^{n-1} \big[k_\perp^m \,\psi(r_\perp)\big]^\dagger 
\big[k_\perp^{\,n-1-m}\,\phi(r_\perp)\big ]\,.
\end{align}
This identity generalises the familiar integration-by-parts formula used for the Weyl-node Hamiltonian. The base case of $n=1$ is immediate, when $\psi^\dagger(k_\perp\phi)-(k_\perp\psi)^\dagger\phi
= -\, i\,\partial_{r_\perp}(\psi^\dagger \, \phi)$, and is exactly the manipulation used implicitly for the linear-in-$k_\perp$ Weyl Hamiltonian. Assuming Eq.~\eqref{eqlagrange} holds up to degree $(n-1)$, writing $k_\perp^n\phi=k_\perp \,(k_\perp^{n-1}\, \phi)$, and adding and subtracting $(k_\perp^{n-1} \,\psi)^\dagger \,(k_\perp \,\phi)$ reduces the degree-$n$ identity to the base case (applied to the pair, $\psi$ and $ k_\perp^{n-1} \, \phi$) plus the degree-$(n-1)$ identity [applied to the pair $k_\perp \, \psi$ and $\phi$]. This completes the induction and fixes the coefficients on the right-hand side of Eq.~\eqref{eqlagrange} uniquely.

\item  Step 3 (boundary functional): 
\\Multiplying Eq.~\eqref{eqlagrange} by the constant (i.e., $r_\perp$-independent) Hermitian matrix $C_n$, summing over $n$, and integrating over the half-space gives
\begin{align}
\label{eqxi}
\psi^\dagger(r_\perp)\, H_{r_\perp}\, \phi(r_\perp) 
- \big[ H_{r_\perp}\psi(r_\perp)\big ]^\dagger\phi(r_\perp)
= -\, i\,\partial_{r_\perp}\, \Xi[\psi,\phi](r_\perp)\,, \quad
\Xi[\psi,\phi] \equiv \sum_n \sum_{m=0}^{n-1} \big(k_\perp^m\psi\big)^\dagger\, C_n\, \big(k_\perp^{\,n-1-m}\phi\big)\,.
\end{align}
Integrating over $r_\perp\in[0,\infty)$, the contribution at $r_\perp\to\infty$ vanishes because $\psi,\phi \,\propto e^{-\kappa \, r_\perp}$ [with $\mathrm{Re}(\kappa)>0$] decay exponentially together with all their derivatives, so that Eq.~\eqref{eqherm1} becomes simply $\Xi[\psi,\phi](0)=0$.

\item  Step 4 (physical boundary state): 
\\It remains to evaluate $\Xi[\psi,\phi](0)$ on the actual candidate surface state, $\psi(r_\perp)=\phi(r_\perp)=\psi_0\, e^{-\kappa \, r_\perp}$. Writing  $\lambda \equiv i\, \kappa$, so that $k_\perp^m\, \psi_0=\lambda^m \, \psi_0$ for every $m$ (since $-\,i \, \partial_{r_\perp} e^{-\kappa \, r_\perp}=i\, \kappa\, e^{-\kappa \, r_\perp}$). Here one must be careful: normalisability of the surface state fixes only $\kappa_r\equiv\mathrm{Re}(\kappa)>0$, while $\kappa_i \equiv\mathrm{Im}(\kappa)$ is, in general, an independent real unknown. The latter us solved-for jointly with $E$ [cf. the parametrisation below Eq.~\eqref{eqhambdy}]. Consequently, $\mathrm{Im}(\lambda)=\kappa_r>0$ strictly, implying $\lambda$ is never real, and $(k_\perp^m\psi_0)^\dagger=\bar\lambda^{\,m} \, \psi_0^\dagger$, with $\bar\lambda\ne\lambda$ in general. The $m$-sum defining $\Xi$ must therefore be summed as the (generally non-degenerate) finite geometric series, leading to
\begin{align}
\sum_{m=0}^{n-1}\bar\lambda^{\,m} \,\lambda^{\,n-1-m}
= \frac{\lambda^n-\bar\lambda^{\,n}}{\lambda-\bar\lambda}\,,\quad \lambda\ne\bar\lambda\,,
\end{align}
rather than collapsing termwise to $n \,\lambda^{n-1}$. This gives
\begin{align}
\Xi[\psi_0,\psi_0](0)=\sum_n \psi_0^\dagger \, C_n \,\psi_0\,
\frac{\lambda^n-\bar\lambda^{\,n}}{\lambda-\bar\lambda}
=\frac{\psi_0^\dagger \,H_N(\lambda)\,\psi_0-\psi_0^\dagger \, H_N(\bar\lambda)\,\psi_0}{\lambda-\bar\lambda}\,,
\end{align}
where $H_N(\lambda)\equiv \sum_n C_n \, \lambda^n$ denotes the matrix polynomial $H_N(k_\perp,\boldsymbol k_\parallel)$, with $k_\perp\to\lambda$ (a c-number substitution, not an operator one). Now, $\psi_0$ solves the eigenvalue equation of Eq.~\eqref{eqhambdy} specialised to this ansatz, $H_N(\lambda)\, \psi_0=E \, \psi_0$ --- hence, $\psi_0^\dagger \,H_N(\lambda) \,\psi_0=E\,|\psi_0|^2$.
Since $C_n^\dagger=C_n$, we have $H_N(\bar\lambda)=H_N(\lambda)^\dagger$, leading to $\psi_0^\dagger\, H_N(\bar\lambda)\,\psi_0
=\big[H_N(\lambda)\psi_0\big]^\dagger \, \psi_0 =E^*\, |\psi_0|^2$. Using $\lambda-\bar\lambda=2 \, i\,\mathrm{Im}(\lambda)=2 \,i \,\kappa_r$, we obtain the exact identity,
\begin{align}
\Xi[\psi_0,\psi_0]\big\vert_{r_\perp=0}=\frac{\big(E-E^*\big)\,|\psi_0|^2}{2\, i\, \kappa_r}
=\frac{\mathrm{Im}(E)}{\kappa_r}\,|\psi_0|^2\,.
\end{align}
Eq.~\eqref{eqherm1} (i.e., $\Xi(0)=0$, from Step~3) is therefore equivalent, order by order, and irrespective of the polynomial degree of $H_N$ in $k_\perp$, to the single real condition, $\mathrm{Im}(E)=0$. The Hermiticity/hard-wall requirement is precisely the statement that a genuine surface-bound state must carry a real energy eigenvalue, as any state in the domain of a self-adjoint operator must. This is the general form of the boundary condition within the class of polynomial continuum Hamiltonians considered here, and is the condition actually enforced [as one real equation among the $(2N+1)$ counted below Eq.~\eqref{eqhambdy}]. It reduces to the familiar statement of Eq.~\eqref{eqcur}, $\psi_0^\dagger \, j_{r_\perp}  \psi_0 \big \vert_{r_\perp=0}=0$, precisely when $H_N$ is at most linear in $k_\perp$ [i.e., $H_N=C_1 \,k_\perp+C_0(\boldsymbol k_\parallel)$, as for the untilted WSM]. There, $j_{r_\perp}=\partial_{k_\perp}H_N=C_1$ is $\kappa$-independent and Hermitian, the $m$-sum contains the single term $m=0$ (so no $\lambda,\bar\lambda$ dependence enters at all], and $\Xi=\psi_0^\dagger \, C_1  \,  \psi_0$ is already manifestly real -- consistent with, and equal to, $\mathrm{Im}(E)|\psi_0|^2/\kappa_r$ above.

The same result also follows directly from the underlying physical continuity equation, independently of the geometric-sum bookkeeping above. Since $\psi(r_\perp)$ solves $H_{r_\perp} \, \psi(r_\perp) = E \, \psi(r_\perp)$ pointwise, we have $\psi^\dagger(r_\perp) \, H_{r_\perp} \, \psi(r_\perp) = E \, |\psi(r_\perp)|^2$ and $\big(H_{r_\perp} \, \psi(r_\perp)\big)^\dagger \, \psi(r_\perp) = E^* \, |\psi(r_\perp)|^2$. Thus, the Step~3 identity, specialised to $\phi=\psi$, reads $\big(E-E^*\big) \, |\psi(r_\perp)|^2 = -i \, \partial_{r_\perp} \, \Xi[\psi,\psi](r_\perp)$, i.e., $\partial_{r_\perp} \, \Xi[\psi,\psi](r_\perp) = -2 \, \mathrm{Im}(E) \, |\psi(r_\perp)|^2$. Integrating from $r_\perp=0$ to $\infty$, using $\Xi[\psi,\psi](\infty)=0$ by decay and $\int_0^\infty |\psi_0|^2 \, e^{-2 \, \kappa_r \, r_\perp} \, dr_\perp = |\psi_0|^2/(2 \, \kappa_r)$, reproduces $\Xi[\psi_0,\psi_0](0) = \mathrm{Im}(E) \, |\psi_0|^2/\kappa_r$ exactly, for any polynomial degree of $H_N$. We emphasise that this is not the same statement as $\mathrm{Re}\big[\psi_0^\dagger \, j_{r_\perp} \, \psi_0\big]=0$: the operator $j_{r_\perp} = \partial_{k_\perp} H_N$ is Hermitian only when evaluated at real $k_\perp$, whereas at the boundary it is evaluated at the complex value $k_\perp\to\lambda$. Therefore, $\psi_0^\dagger \, j_{r_\perp} \, \psi_0\big\vert_{r_\perp=0}$ is an analytic continuation rather than a genuine Hermitian expectation value, and need not coincide with the manifestly real $\Xi[\psi_0,\psi_0](0)$ beyond quadratic order in $k_\perp$, as we illustrate explicitly below.

\end{itemize}

As an explicit illustration beyond the Weyl case, consider the double-Weyl Hamiltonian of Eq.~\eqref{eqhamdw}, for which $C_0$, $C_1$, and $C_2$ are all nonzero, so that $H_N$ is quadratic in $k_\perp=k_x$. For any polynomial containing only powers up to $n=2$, the geometric sum of Step~4 simplifies further: since the $n=2$ factor obeys $(\lambda^2-\bar\lambda^2)/(\lambda-\bar\lambda)=\lambda+\bar\lambda=2\,\mathrm{Re}(\lambda)$ for any complex $\lambda$, the $n=2$ contribution to $\Xi[\psi_0,\psi_0](0)$ reduces to $2\,\mathrm{Re}(\lambda)\,\psi_0^\dagger \,  C_2 \, \psi_0$. Hence, it coincides exactly with the real part of the corresponding term of $\psi_0^\dagger \,j_{r_\perp}\, \psi_0$. Combined with the $n\le1$ terms, which are already manifestly real, this gives $\Xi[\psi_0,\psi_0](0)=\mathrm{Re}\big[\psi_0^\dagger  \, j_{r_\perp} \! \psi_0\big]_{r_\perp=0}$, whenever $H_N$ is at most quadratic in $k_\perp$. Thus, $\mathrm{Im}(E)=0$ is equivalent, for such Hamiltonians, to the real part of the current condition of Eq.~\eqref{eqcur}, even though $\kappa$ itself remains complex. Using $\psi_0^T=[1 \quad a_1+i\,b_1]$ and $j_x=\partial_{k_x}H_{dW}=2 \, k_x\,\sigma_x+2 \, k_y\,\sigma_y$ (with $k_0$ set to $1$), one finds $\psi_0^\dagger\,\sigma_x\,\psi_0=2\, a_1$ and $\psi_0^\dagger \, \sigma_y \, \psi_0=2 \,b_1$. This leads to $\psi_0^\dagger  \, j_x \! \psi_0\big\vert_{x=0}=2 \, i \, \kappa \, (2 \, a_1)+2 \, k_y \, (2 \, b_1)=(4 \, b_1 \, k_y-4 \, a_1 \, \kappa_i)+4 \, i \, a_1 \, \kappa_r$, whose real part vanishing gives exactly $\kappa_i=(b_1/a_1)\,k_y$ [reproducing Eq.~\eqref{eq29}]. This confirms explicitly, for the case in the paper where $H_N$ is quadratic, that the general condition $\mathrm{Im}(E)=0$ derived above reduces to the boundary condition already used in Sec.~\ref{sec-dws}. For Hamiltonians containing $k_\perp^n$ with $n\ge3$ (e.g., the triple- and quadruple-Weyl models), this intermediate identification with $\mathrm{Re}\big[\psi_0^\dagger  \, j_{r_\perp}\! \psi_0\big]$ need not hold term by term. There, $\mathrm{Im}(E)=0$ itself, established generally in Step~4 above, remains the quintessential boundary condition.
}}

\bibliography{ref_fa}
  
\end{document}